\documentclass[12pt,twoside,a4paper]{article}  

\usepackage{amssymb,amsfonts,amsmath,amsbsy,theorem,amscd, dina42e}

\def\bsymbol#1{\mbox{\boldmath$\displaystyle#1$\unboldmath}}

\newcommand{\bfgamma}{{\bsymbol{\gamma}}}
\newcommand{\bfnu}{{\bsymbol{\nu}}}
\newcommand{\bftau}{{\bsymbol{\tau}}}

\numberwithin{equation}{section}

\newcommand{\R}{\ensuremath{\mathbb{R}}}

\newcommand{\nabs}{\nabla_{\!s}}

\newcommand{\EXP}{\operatorname{exp}}

\newcommand{\sd}{\, d}

\newcommand{\eps}{\ensuremath{\varepsilon}}

\newcommand{\Div}{\operatorname{div}}

\newcommand{\habil}[1]{}
\newcommand\J{{\mathbf J_\varphi}}
\newcommand{\Jw}{{\mathbf J_w}}
\newcommand\bu{{\mathbf v}}
\newcommand\bs{{ s}}
\newcommand\K{{\mathbf K}}
\newcommand{\stress}{{\mathbf S}}
\newtheorem{thm}{THEOREM}[section]

\newtheorem{lem}[thm]{Lemma}

\newtheorem{theorem}[thm]{Theorem}

\newtheorem{claim*}{Claim}

\theorembodyfont{\rmfamily}
\newtheorem{rem}[thm]{Remark}

\newenvironment{proof*}[1]{{\bf Proof #1:}}{\hspace*{\fill}\rule{1.2ex}{1.2ex}\\ } 
\newenvironment{proof}{{\bf Proof:\,}}{\hspace*{\fill}\rule{1.2ex}{1.2ex}\\ }

\newcommand{\hrho}{\hat{\rho}}

\newcommand{\dichtabl}{{\frac{\partial\rho}{\partial\varphi}}}
\newcommand{\ve}{\mathbf{v}} 
\newcommand{\tn}[1]{\mathbf{#1}}
\newcommand{\vc}[1]{\mathbf{#1}}
\newcommand{\dsm}{\sd s_x} 
\newcommand{\nl}{\boldsymbol{\nu}} 
\newcommand{\Om}{{\Omega}}
\begin{document}
\begin{titlepage}
\title{Thermodynamically Consistent, Frame Indifferent Diffuse Interface Models for
  Incompressible Two-Phase Flows with Different Densities
\footnote{This preprint is
a modified version of the preprint~\cite{ModifiedModelH1}.}
}
\author{  Helmut Abels\footnote{Fakult\"at f\"ur Mathematik,  
Universit\"at Regensburg,
93040 Regensburg,
Germany, e-mail: {\sf helmut.abels@mathematik.uni-regensburg.de}}, Harald Garcke\footnote{Fakult\"at f\"ur Mathematik,  
Universit\"at Regensburg,
93040 Regensburg,
Germany, e-mail: {\sf harald.garcke@mathematik.uni-regensburg.de}}, and G\"unther Gr\"un\footnote{Department Mathematik,
Universit\"at Erlangen-N\"urnberg,
Martensstr. 3,
91058 Erlangen, 
Germany, e-mail: {\sf gruen@am.uni-erlangen.de}}}
\date{}
\end{titlepage}
\maketitle
\begin{abstract}
A new diffuse interface model for a two-phase  flow of two incompressible
fluids with different densities is introduced using methods from rational
continuum mechanics. The model fulfills local and global dissipation inequalities and is frame indifferent. Moreover, it
 is generalized to situations with a soluble species.
Using the method of matched asymptotic expansions we derive various sharp interface models
in the limit when the interfacial thickness tends to zero. Depending on the scaling 
of the mobility in the diffusion equation we either derive classical sharp interface models or
models where bulk or surface diffusion is possible in the limit. In the latter case a new term resulting from surface diffusion appears in the momentum balance at the interface. Finally, we
show that all sharp interface models fulfill natural energy inequalities.
\end{abstract}

\noindent{\bf AMS-Classification:} 
Primary: 76T99; Secondary:
35Q30, 
35Q35, 
35R35,
76D05, 
76D45, 
80A22

\section{Introduction}

In recent years diffuse interface models have been successfully used  to describe the flow of two or more
immiscible fluids both for theoretical studies and numerical simulations. One fundamental advantage of these
models is that they are able to describe topological transitions like droplet coalescence or droplet break-up in
a natural way. These ideas go even back to works of van der Waals and Korteweg, cf.~\cite{DiffIntModels} for a review and further references.  In the case of two incompressible, viscous Newtonian fluids the basic diffuse interface model
is the so-called ``Model H'', cf. Hohenberg and
Halperin~\cite{HohenbergHalperin}. It leads to the 
Navier-Stokes/Cahn-Hilliard system
\begin{alignat}{2}\label{eq:NSCH1}
  \rho\partial_t \ve + \rho(\ve\cdot \nabla) \ve - \Div (2\eta (c)D\ve) +
  \nabla p &= -\hat\sigma\eps\Div (\nabla c\otimes 
\nabla c),
\\\label{eq:NSCH2}
  \Div \ve &=0, 
\\\label{eq:NSCH3}
  \partial_t c + \ve\cdot\nabla c &= \Div (m\nabla \mu), 
\\\label{eq:NSCH4}
\mu &= \hat\sigma\eps^{-1}\psi'(c) - \hat\sigma\eps\Delta  c. 
\end{alignat}
Here $\rho$ is the density, $\ve$ is the mean velocity, $D\ve=\frac12(\nabla \ve+\nabla \ve^T)$, $p$ is the pressure, and $c$ is an
order parameter related to the concentration of the fluids (e.g. the concentration difference or the
concentration of one component). Moreover, $\eta(c)>0$ is the viscosity of the
mixture, $\hat\sigma$ is a constant related to the surface energy density, $\eps>0$ is a (small) 
parameter, which is related to the ``thickness'' of the interfacial region, 
 $\psi$ is a homogeneous free energy
density and $\mu$ is the chemical potential. Capillary forces due to surface tension are modeled by an extra contribution $\eps \nabla c \otimes
\nabla c:= \eps \nabla c (\nabla c)^T$ in the stress tensor leading to the term on the right-hand side of (\ref{eq:NSCH1}). Moreover, we note
that in the modeling diffusion of the fluid components is taken into account. Therefore $\Div (m\nabla \mu)$ is
appearing in (\ref{eq:NSCH3}), where $m=m(c)\geq 0$ is the mobility coefficient.

One of the fundamental modelling assumptions is that the densities of both components  as well as the density of
the mixture $\rho$ are constant. Of course, this restricts the applicability of the model to situations when density differences are negligible.
Gurtin et al.~\cite{GurtinTwoPhase} derived this model in the framework of rational continuum mechanics and
showed that it satisfies the second law of thermodynamics in a mechanical version based on a local dissipation
inequality.

Lowengrub and Truskinovsky~\cite{LowengrubQuasiIncompressible} derived a thermodynamically consistent extension
of the Model H for the case of different densities, which leads to the system:
\begin{alignat}{2}\label{eq:LT1}
  \rho \partial_t \ve + \rho (\ve\cdot \nabla) \ve -\Div \mathbf S(c,D\ve)
  + \nabla p&= -\hat\sigma\eps\Div (\rho \nabla c\otimes \nabla c),\\
  \label{eq:LT2}
  \partial_t \rho +\Div (\rho \ve)&= 0, \\
  \label{eq:LT3}
  \rho \partial_t c+ \rho \ve\cdot \nabla c &= \Div(m(c)\nabla \mu),
\\  \label{eq:LT4}
\mu &= -\rho^{-2}\tfrac{\partial \rho}{\partial c} p+
\tfrac{\hat\sigma}\eps\psi'(c) - \tfrac{\hat\sigma\eps}{\rho}\Div (\rho\nabla c),
\end{alignat}
where $\tn{S}(c,D\ve)=2\eta(c)D\ve +\lambda(c) \Div \ve\, \tn{I}$ and $\lambda(c)$ is the bulk viscosity coefficient.
Here the free energy has the density $\rho \hat\sigma(\eps^{-1}\psi (c) +\eps \frac{|\nabla c|^2}2)$ per unit volume.
A simplified version of this model has been successfully used for numerical studies, cf. Lee et
al.~\cite{LeeLowengrub1,LeeLowengrub2}. 
In contrast, there are -- to the best of the authors' knowledge -- no discrete
schemes available which are based on the full model
\eqref{eq:LT1}-\eqref{eq:LT4}. This may be due to fundamental new difficulties
compared with Model H \eqref{eq:NSCH1}-\eqref{eq:NSCH4}.
  For instance,  
the velocity field $\ve$ is no longer divergence free and 
therefore no solution concept is available
which avoids to   determine the pressure $p$.
At least analytically, these difficulties could be overcome, see
Abels~\cite{LTModel} for existence of weak solutions. Mathematically the
coupling of the Navier-Stokes \eqref{eq:LT1}-\eqref{eq:LT2} and the
Cahn-Hilliard part 
\eqref{eq:LT3}-\eqref{eq:LT4} is much stronger, for instance 
the pressure $p$
 enters the equation for the chemical potential \eqref{eq:LT4},
 and the linearized system is
very different from the 
linearization of  Model H, cf. Abels~\cite{LTModelShortTime}, where strong
solutions locally in time are constructed.

Alternative generalizations of the Model H for the case of different densities were presented and discussed by
Boyer~\cite{BoyerModel} and Ding et al.~\cite{DingSpeltShu}. The model by Ding et al. consists of~\eqref{eq:NSCH1}-\eqref{eq:NSCH4}, but now for a variable
density $\rho=\rho(c)$. In order to justify this generalization they start from the mass balance equation
\begin{equation}\label{eq:MassBalance}
  \partial_t \rho_j +\Div (\rho_j \ve_j)=0
\end{equation}
for the individual fluids $j=1,2$ and define the mean velocity $\ve$ of the mixture as \emph{volume averaged
velocity} $\ve= u_1 \ve_1+u_2 \ve_2$, where $u_j$ is the volume fraction of fluid $j$. Then \eqref{eq:MassBalance}
yields
\begin{equation}\label{eq:DivFree'}
  \Div \ve =0,
\end{equation}
cf. Section~\ref{eq:Derivation} below.
In contrast to that Lowengrub and Truskinovsky define the mean velocity $\ve$ as \emph{mass averaged/barycentric
velocity} $\rho\ve= \rho_1\ve_1+\rho_2 \ve_2$, which yields
\begin{equation*}
  \partial_t \rho +\Div(\rho \ve) =0.
\end{equation*}
The incompressibility relation \eqref{eq:DivFree'}  of course has advantages with respect to numerical simulations -- see the computations related to the model by Ding et al. in \cite{DingSpeltShu}. 

Unfortunately, neither global nor local energy inequalities are known to hold for
\eqref{eq:NSCH1}-\eqref{eq:NSCH4}  when $\rho$ is
not constant. The model by Boyer is more
complicated. But it is derived using a volume averaged mean velocity, which leads
to a divergence free mean velocity field too. The further derivation of Boyer differs from the one in \cite{DingSpeltShu} and ours since the starting point are the equations for the conservation of linear momentum of each single fluid. Also for this model neither global nor local energy inequalities seem to be known, cf. also \cite{BoyerNonMatched}.  
\newline  It is the purpose of the present paper to derive a
thermodynamically consistent generalization of \eqref{eq:NSCH1}-\eqref{eq:NSCH4} to the case of non-matched
densities based on a solenoidal velocity field $\ve$. More precisely, we will derive the system
\begin{alignat*}{2}
   \partial_t(\rho \ve) + \Div (\rho \ve\otimes \ve) &+ \Div (\ve\otimes \tfrac{\tilde{\rho}_1-\tilde{\rho}_2}2 m(\varphi)\nabla\mu) -\Div (2\eta(\varphi)D\ve)+ \nabla p\\
&= -\hat{\sigma}\eps\Div ( \nabla
\varphi \otimes \nabla \varphi),\\
  \Div \ve&= 0, \\
  \partial_t \varphi+ \ve\cdot \nabla \varphi  &= \Div(m(\varphi) \nabla \mu_\varphi),
\end{alignat*}
together with
\begin{equation*}
 \mu_\varphi = \hat{\sigma}\eps^{-1}\psi^\prime (\varphi) -\hat{\sigma}\eps\Delta\varphi,
\end{equation*}
where the order parameter $\varphi=\varphi_2-\varphi_1$ stands for the difference of 
the volume fractions $\varphi_j$, $j=1,2$\footnote{Other choices of order parameters are possible as well -- see Section~\ref{eq:Derivation}.}.
Here the free energy of the system has the density $\hat{\sigma}\eps^{-1}f(\varphi)+\hat{\sigma}\eps\frac{|\nabla \varphi|^2}2$ (per
unit volume).  We note that the first equation in the latter system is equivalent to
\begin{equation*}
     \rho\partial_t \ve + \left(\left(\rho \ve+\tilde{\mathbf{J}}\right)\cdot \nabla
\right) \ve  -\Div (2\eta(\varphi)D\ve)+ \nabla p = -\hat{\sigma}\eps\Div ( \nabla
\varphi \otimes \nabla \varphi),
\end{equation*}
where $\tilde{\mathbf{J}}=\tfrac{\tilde{\rho}_1-\tilde{\rho}_2}2 m(\varphi)\nabla\mu$, cf. Section~\ref{eq:Derivation} below.
In comparison with the system derived in \cite{DingSpeltShu} there is the additional term $(\tilde{\mathbf{J}}\cdot \nabla) \ve$ in the equation for the linear momentum. This term
vanishes in the case of matched densities, i.e., $\rho\equiv const.$. But this term is crucial in the case of
non-matched densities for consistency with thermodynamics.
We note that in contrast to the model by Lowengrub and Truskinovsky \eqref{eq:LT1}-\eqref{eq:LT4} the usual
continuity equation \eqref{eq:LT2} is not part of our system. Nevertheless there is conservation of mass in
our system. More precisely, we have
\begin{equation}\label{eq:masscons}
  \partial_t \rho + \Div \left(\rho \ve +\tilde{\mathbf{J}} \right)=0,
\end{equation}
and in Section~\ref{eq:Derivation} it will be shown that in fact individual masses are conserved. 
Note that $\tilde{\rho}_j$ is the specific density of fluid $j=1,2$ and that $\rho=
\frac{\tilde{\rho}_1+\tilde{\rho}_2}2+\frac{\tilde{\rho}_2-\tilde{\rho}_1}2\varphi$.
We emphasize that according to equation~\eqref{eq:masscons} the (volume averaged) velocity $\ve$ does not describe the flux of the density. In our
model the flux of the density consists of the two parts: $\rho \ve$, describing the transport by the mean
velocity, and a relative flux $\tilde{\mathbf{J}}=-\frac{\tilde{\rho}_2-\tilde{\rho}_1}2 m(\varphi)\nabla \mu$ related to diffusion
of the components. Hence the diffusion of the components relative to the mean velocity leads to a diffusion of
the mass density in the case that $\tilde{\rho}_1\neq \tilde{\rho}_2$. Moreover, we note that in the classical
Model H effects related to diffusion of the components can play an important role and can lead to Ostwald
ripening effects and disappearance of small droplets, cf. e.g. \cite{SpontaneousShrinkage}.

The structure of the paper is as follows: In Section~\ref{eq:Derivation} we will derive the generalization of
Model H, described above in the framework of rational continuum mechanics. First we will use a local dissipation
inequality and a choice of the energy flux as in \cite{GurtinTwoPhase} to derive restrictions for the form of
the stress tensor and the chemical potential, which finally leads to our model after suitable constitutive
assumptions. Then we briefly discuss the changes in the derivation if the energy flux is not specified at the
beginning and Liu's Lagrange multiplier method is used, cf. \cite{Liu}. In Section~\ref{sec:Onsager}, we  present
a third approach to derive a thermodynamically consistent model, this time based on Onsager's variational
principle. We consider the more general situation when either the system is subjected to gravitational forces or when one additional soluble species is present
in both fluids. In the latter case, transport effects across the interface are taken into account, too,  and we derive a
diffuse interface analogue of Henry's law, cf. Section \ref{sec:Onsager}.

In Section~\ref{sec:SharpInterface} we discuss the sharp interface asymptotics in the limit $\eps\to 0$ for the diffuse
interface model together with a soluble species. This is done by using the method of formally matched
asymptotics. We show that the limit system depends essentially on the choice and the scaling of the mobility.
Actually, we consider four cases related to choosing the mobility degenerate or non-degenerate and letting the
mobility tend to zero or not. If the mobility $m(\varphi)$ vanishes as $\eps\to 0$, we end up with the classical model
for a two-phase flow with the Young-Laplace law
\begin{equation*}
  - [2\eta D\ve]_-^+\boldsymbol{\nu}+ [p]_-^+\boldsymbol{\nu} = \sigma \kappa \boldsymbol{\nu}\qquad \text{at}\
    \Gamma(t),
\end{equation*}
where $\Gamma(t)$ is the interface between the fluids, $\boldsymbol{\nu}$ is a unit normal to $\Gamma(t)$, $\kappa$ is its mean
curvature, $\sigma$ is a surface tension coefficient, and $[.]_-^+$ denotes the jump of a quantity at $\Gamma(t)$ in the direction of $\boldsymbol{\nu}$. Moreover, the interface is transported by the velocity of the fluid, i.e.,
\begin{equation*}
  \mathcal{V}- \boldsymbol{\nu}\cdot \ve=0 \qquad \text{at}\ \Gamma(t),
\end{equation*}
where $\mathcal{V}$ is the normal velocity of $\Gamma(t)$. If a soluble species
with density $w$  is present, we obtain the classical Henry condition for the
jump of the concentrations of the soluble species.

In the case of a constant mobility, we obtain in the limit $\eps\to 0$ that the momentum balance in the bulk is given as
\begin{equation*}
  \tilde{\rho}_i\partial_t\ve +\Div \left(\ve\otimes\left(\tilde{\rho}_i\ve + \tfrac{\tilde{\rho}_1-\tilde{\rho}_2}2 m_0 \nabla \mu\right)\right)-\eta_i\Delta \ve +\nabla p=0. 
\end{equation*}
In this case the diffusive flux $-\tfrac{\tilde{\rho}_2-\tilde{\rho}_1}2 m_0 \nabla \mu$ enters the momentum balance, cf. \cite{AltEntropy}.
At the interface we obtain the Stefan type condition
\begin{eqnarray*}
2(\bu\cdot\bfnu-\mathcal{V}) &=& m_0[\nabla \mu]^+_-\cdot\bfnu\qquad \text{at}\
    \Gamma(t),
\end{eqnarray*}
for the evolution of the interface and again
\begin{eqnarray*}
-[2\eta D\bu]^+_-\bfnu+[p]^+_-\bfnu
 &=& \sigma\kappa\bfnu \qquad \text{at}\
    \Gamma(t),
\end{eqnarray*}
for the jump of the stress tensor.
Here  $\mu$ satisfies
\begin{eqnarray*}
2\mu &=&\sigma\kappa-[w]^+_-\qquad \text{at}\ \Gamma(t),
\end{eqnarray*}
$\mu$ is harmonic in the bulk, and $m_0>0$ is a diffusion coefficient related to $m$.
In particular the interface is no longer material and 
 diffusion of mass 
through the bulk is still present in the model.
In the case of a non-vanishing, degenerate mobility the evolution of the interface is governed by the surface diffusion law
\begin{equation*}
2(\bu\cdot\bfnu-\mathcal{V}) = \hat{m}\Delta_{\Gamma(t)}\mu,
\end{equation*}
where $\Delta_{\Gamma(t)}$ is the Laplace-Beltrami operator of $\Gamma(t)$ and  $\hat{m}>0$ is a diffusion coefficient related to $m$. Moreover, the equation for the jump of the stress tensor is
\begin{equation*}
-\tfrac{\tilde{\rho}_2-\tilde{\rho}_1}2\hat{m}((\nabla_\Gamma\mu)\cdot\nabla_\Gamma)\ve-[2\eta D\bu]^+_-\bfnu+[p]^+_-\bfnu
 = \sigma\kappa\bfnu \qquad \text{at}\
    \Gamma(t),
\end{equation*}
in this case. In particular the surface flux $-\tfrac{\tilde{\rho}_2-\tilde{\rho}_1}2\hat{m}\nabla_\Gamma\mu$ enters the momentum balance at the interface.
In Section \ref{sec:Energy} we prove that  energy estimates are valid for sufficiently smooth solutions of
the sharp interface models.
Finally, several important identities for the formally matched asymptotics
calculations are shown in the appendix.

\medskip

\noindent
{\bf Acknowledgements:} The authors are grateful to  Hans Wilhelm Alt for several stimulating
discussions and the telling us about \cite{AltEntropy}, which resulted in the present
model and helped to improve the presentation of the derivation of the model. Moreover,
we thank the anonymous referees for their valuable comments on the previous version of
this paper.
This work was supported by the SPP 1506 "Transport Processes at Fluidic Interfaces" of
the German Science Foundation (DFG) through the grants GA 695/6-1 and 
GR 1693/5-1.

\section{Derivation of the Model}\label{eq:Derivation}
In order to derive the diffuse interface model, one assumes a partial mixing of the macroscopically immiscible fluids in a thin interfacial region.  Therefore let us introduce some terminology related to mixtures. In the following the fluids are labeled by $j=1,2$ and  they fill a domain $\Omega\subseteq\R^d$.
The total mass density of the mixture is denoted by $\rho$. Moreover, $\rho_j$ denotes the mass density of the fluid $j$, i.e.,
\begin{equation*}
  M_j= \int_V \rho_j(x) \sd x
\end{equation*}
is the mass of the fluid $j$ contained in a set $V\subset \Omega$ and we obtain $\rho=\rho_1+\rho_2$. 
Moreover, we denote by $c_j= \frac{\rho_j}\rho$ the \emph{mass concentration} 
and note that $c_1+c_2=1$.

Denoting by $\hat {\vc{J}}_j $ the mass flux of fluid $j$, the mass balance equation in local form is given
by
\begin{equation*}
 \partial_t \rho_j + \Div \hat{ \vc{J}}_j=0.
\end{equation*} 
Defining the velocities $\mathbf v_j$, $j=1,2$, of the single fluids as
$\mathbf v_j = \hat {\vc{J}}_j/\rho_j$ the mass balance equation can be rewritten as
$$ \partial_t \rho_j + \Div (\rho_j \mathbf v_j ) =0.$$
In what follows we assume that the volume occupied by a given amount of mass of the 
single fluids does not change after mixing, i.e., the excess volume due to
mixing is zero. If
${\tilde{\rho}_j}$ is the specific  (constant) density of the unmixed fluid $j$, we introduce
$u_j= \frac{\rho_j}{\tilde{\rho}_j}$. The assumption that the excess volume is zero results in
\begin{equation}
\label{volin}
u_1+u_2=1.
\end{equation}
Expressed in terms of the mass concentrations $c_1$ and $c_2$, condition (\ref{volin})
reads as
\begin{equation*}
\frac{c_1 \, \rho}{\tilde{\rho}_1} +\frac{c_2 \, \rho}{\tilde{\rho}_2} =1 \qquad \Longleftrightarrow \qquad
\frac 1\rho = \frac{c_1 }{\tilde{\rho}_1} +\frac{c_2 }{\tilde{\rho}_2}.
\end{equation*}
Introducing the mass concentration difference $c=c_2-c_1$, the above relation  implies that
$\rho=\hat \rho (c) $ with a function $\hrho$ defined via
\begin{equation*}
 \frac1{\hrho(c)}=  \frac{\frac12(c+1)}{\tilde{\rho}_2} + \frac{\frac12(1-c)}{\tilde{\rho}_1}.
\end{equation*}
We remark that possible choices for the order parameter in the phase field model are the 
mass concentration difference $c$, the density difference $\bar \rho:=\rho_2-\rho_1$ or the difference of volume
fractions $u:=u_2-u_1.$

We now introduce a suitable averaged velocity of the mixture. In 
contrast to the
 \emph{mass averaged/barycentric velocity} $\tilde{\ve}$ given by $\rho \tilde{\ve}= \rho_1 \ve_1+\rho_2 \ve_2$, cf. Lowengrub and
Truskinovsky~\cite{LowengrubQuasiIncompressible}, we choose the volume averaged velocity $\ve$ of the mixture as in Boyer~\cite{BoyerModel} and Ding et al.~\cite{DingSpeltShu}.
More precisely, 
 we define
 \begin{equation*}
   \ve= u_1 \ve_1 + u_2 \ve_2 =\frac {\rho_1}{\tilde \rho_1} \ve_1+
\frac {\rho_2}{\tilde \rho_2} \ve_2 ,
 \end{equation*}
cf. \cite{BoyerModel,DingSpeltShu}.
As a consequence, we obtain, using the fact that the ${\tilde{\rho_j}}$'s are constants,
 \begin{equation}\label{eq:DivFree}
   \Div \ve = \Div (\tfrac{\rho_1}{\tilde{\rho_1}}\ve_1)+ \Div (\tfrac{\rho_2}{\tilde{\rho_2}}\ve_2)=
\Div (\tfrac{ \hat {\vc{J}}_1} {\tilde\rho_1}+   \tfrac{ \hat {\vc{J}}_2} {\tilde\rho_2} ) =
- \partial_t
\left(\tfrac{\rho_1}{\tilde{\rho_1}}+\tfrac{\rho_2}{\tilde{\rho_2}} \right)= -\partial_t 1 =0
.
 \end{equation}
Furthermore, we denote by  $\vc{J}_j=\hat {\vc{J}}_j-\rho_j \ve$ the mass flux of the fluid $j$ relative to the velocity $\ve$, i.e.,
\begin{equation*}
  \partial_t \rho_j + \Div (\rho_j \ve) + \Div \vc{J}_j=0.
\end{equation*}
Because of (\ref{eq:DivFree}) and $\rho_j = \rho c_j=\tilde{\rho}_j u_j$, we have
\begin{equation*}
  \partial_t (\rho c_j) + \ve \cdot \nabla (\rho c_j) + \Div \vc{J}_j =0
\end{equation*}
or equivalently,
\begin{equation*}
  \partial_t u_j + \ve \cdot \nabla u_j + \Div \tilde{\vc{J}}_j=0,
\end{equation*}
where $\tilde{\vc{J}}_j= \frac{\vc{J}_j}{\tilde{\rho}_j}$. Because of $u_1+u_2=1$, we require
\begin{equation}
  \label{fluxconstr}
\tilde{\vc{J}}_1+
\tilde{\vc{J}}_2= \frac{\vc{J}_1}{\tilde{\rho}_1}+\frac{\vc{J}_2}{\tilde{\rho}_2}=0,
\end{equation}
 cf. (\ref{eq:DivFree}).
In particular, we obtain
\begin{equation}\label{eq:Flux}
  \partial_t (\rho c) + \ve\cdot \nabla (\rho c) +\Div \vc{J}=0,
\end{equation}
where $\vc{J}=\vc{J}_2-\vc{J}_1$.
In addition we have
\begin{equation}\label{mass1}
\partial_t \rho = \partial_t (\rho_1+\rho_2) = -\Div( {\vc{J}}_1 +\rho_1 \ve +{\vc{J}}_2 +\rho_2\ve)
= -\Div( \rho \ve + {\vc{J}}_1+ {\vc{J}}_2 ).
\end{equation}
If $\tilde \rho_1 \neq \tilde \rho_2$, we have in general
$ \Div({\vc{J}}_1+ {\vc{J}}_2 ) \neq 0 $. Hence the classical continuity equation does not hold
with respect to the velocity $\ve$. This reflects the fact that we allow for mass diffusion in the system.

Instead of $c$, one could use $u:=u_2-u_1$ as order-parameter. Because of
\begin{gather}
\label{eq:neu}
\rho =\rho c_1+\rho c_2= \tilde{\rho}_1 u_1+\tilde{\rho}_2 u_2, \\
\label{eq:neu2}\rho c=\rho c_2-\rho c_1= \tilde{\rho}_2 u_2-\tilde{\rho}_1 u_1
\end{gather}
and $u_2=\frac{1+u}2$, $u_1=\frac{1-u}2$, we can also assume that $\rho=\hrho(u)$ and $\rho c= \widehat{\rho
c}(u)$. In order to have flexibility in the choice of the order parameter, we assume in the following that $\varphi$ is any suitable order parameter such that  $\rho = \hrho(\varphi)$ and $\rho c= \widehat{\rho c}(\varphi)$ for some constitutive functions $\hrho, \widehat{\rho c}$. Here $\rho c=\widehat{\rho c}(\varphi)$ is the density difference. Possible choices are $\varphi=u_1-u_2$ as above, $\varphi=c$, and $\varphi=\rho c$. Then
\begin{equation}\label{eq:DiffEq1}
  \partial_t \widehat{\rho c}(\varphi) + \ve\cdot \nabla \widehat{\rho c}(\varphi)+\Div \vc{J}=0
\end{equation} 
 is equivalent to
\begin{equation}\label{eq:DiffEq1'}
  \frac{\partial (\widehat{\rho c})}{\partial \varphi}\left(\partial_t \varphi + \ve\cdot \nabla \varphi\right)=
-\Div \vc{J}.
\end{equation}

As in Gurtin et al.~\cite{GurtinTwoPhase}, we assume that the inertia and kinetic energy due to the motion of the fluid relative to the gross motion is negligible. Therefore we consider the mixture as a single fluid with velocity $\ve$, 
 which satisfies the law of conservation of linear   momentum of continuum mechanics with respect to the volume averaged velocity.
The   density and the stress tensor are assumed
to depend on  additional internal variables like $\varphi$ and $\nabla \varphi$.
I.e., we assume that
\begin{eqnarray}\label{eq:ConservationLinMomentum}
  \partial_t(\rho \ve) + \Div (\rho \ve\otimes \ve)&=& \Div \tilde{\tn{T}}  
\end{eqnarray}
for a tensor $\tilde{\tn{T}}$, which has to be specified by constitutive assumptions. Here external forces are neglected for simplicity.

The formulation (\ref{eq:ConservationLinMomentum}) can be rewritten as
\begin{equation}\label{mom1}
(\partial_t\rho+\bu\cdot\nabla\rho)\bu+\rho(\partial_t\bu+(\bu\cdot\nabla)\bu)=\Div\tilde{\tn{T}}\,.
\end{equation}
In the above version the tensor $\tilde{\tn{T}}$ cannot be objective,
i.e., an observer change 
\begin{equation*}
(t^\ast,\mathbf{x}^\ast)=(t,\mathbf{a}(t)+\mathbf{Q}(t)\mathbf{x})
\end{equation*}
will not lead to a transformation rule
\begin{equation*}
\tilde{\tn{T}} = \mathbf{Q}\tilde{\tn{T}}^\ast \mathbf{Q}^{\tn{T}}\,,
\end{equation*}
see \cite{AltEntropy} for details. Using (\ref{mass1}) and $\Div\bu= 0$ we
obtain that the system (\ref{mom1}) can be rewritten as
\begin{equation*}
\rho(\partial_t\bu+(\bu\cdot\nabla)\bu)=\Div(\tilde{\tn{T}}+\bu\otimes(\vc{J}_1+\vc{J}_2))-((\vc{J}_1+\vc{J}_2)\cdot\nabla)\bu\,.
\end{equation*}
This system now allows for an objective tensor $\tn{T}=\tilde{\tn{T}}
+\bu\otimes (\vc{J}_1+\vc{J}_2)$. Using the fact that $\bu$ is
divergence free we introduce the unknown pressure $p$ and restate the
momentum equation as
\begin{equation}\label{eq:momentum1}
\rho(\partial_t\bu+(\bu\cdot\nabla)\bu)=\Div \widetilde{\tn{S}}-\nabla p-((\vc{J}_1+\vc{J}_2)\cdot\nabla)\bu
\end{equation}
where $\widetilde{\tn{S}} = \tn{T}+p\tn{I} = \tilde{\tn{T}}
+\bu\otimes(\vc{J}_1+\vc{J}_2)+p\tn{I}$, where $\tn{I}$ is the identity.
\begin{rem} 
  In the following we use the abbreviation 
\begin{equation*}
\widetilde{\vc{J}} = \mathbf{J}_1+\mathbf{J}_2
\end{equation*}
Because of \eqref{fluxconstr}, we have
\begin{equation}\label{eq:RelationJ}
  \widetilde{\vc{J}}= \frac{\tilde{\rho}_2-\tilde{\rho}_1}{\tilde{\rho}_2+\tilde{\rho}_1} \vc{J}.
\end{equation}
Then the total mass balance (\ref{mass1}) can be rewritten as 
\begin{equation}
\label{gg-new}
\partial_t\rho + \Div(\rho\bu + \widetilde{\vc{J}})=0\,.
\end{equation}
Using the mass balance we can restate the momentum balance
(\ref{eq:momentum1}) as 
\begin{equation}\label{eq:momentum2}
\rho\partial_t\bu+((\rho\bu+\widetilde{\vc{J}})\cdot\nabla)\bu = \Div \widetilde{\tn{S}}-\nabla p
\end{equation}
We note that $\rho \ve+\vc{J} = \rho \tilde{\ve}$ if $\tilde{\ve}$ denotes the barycentric velocity defined by $\rho \tilde{\ve}= \rho_1\ve_1+\rho_2\ve_2$. Moreover, \eqref{eq:momentum2} is equivalent to 
\begin{equation*}
  \frac{d}{dt}\int_{V(t)} \rho \ve \sd x= \int_{\partial V(t)}  \mathbf{T}
\nl \sd s_x
\end{equation*}
for any (sufficiently smooth) domain \emph{$V(t)$ that is transported by $\tilde{\ve}$}, i.e., the normal velocity of $\partial V(t)$ coincides with $\nl\cdot \tilde{\ve}$, cf \eqref{eq:transport} below.
If $V(t)$ is instead transported by the volume averaged velocity $\ve$, then \eqref{eq:momentum2} is equivalent to 
\begin{equation}\label{eq:IntegralMomentumBalance}
  \frac{d}{dt}\int_{V(t)} \rho \ve \sd x= \int_{\partial V(t)} \mathbf{T} \nl
\sd s_x- \int_{\partial V(t)}  (\nl\cdot \tilde{\mathbf{J}})\ve   \sd s_x=
\int_{\partial V(t)}  \tilde{\mathbf{T}} \nl \sd s_x.
\end{equation}
\end{rem}
\vskip 3mm

Finally, we assume the relative motion of the fluids to be diffusive, and we introduce a Helmholtz free energy density $f(\varphi,\nabla\varphi)$ (per unit volume).
It will play the role of an interfacial energy for the diffuse interface. 
The total energy in a volume $V$ is then obtained as the sum of the kinetic and the free energy, i.e.,
\begin{equation*}
  E_V(\varphi,\ve) = \int_V \hrho(\varphi) \frac{|\ve|^2}2 \sd x + \int_V  f(\varphi,\nabla \varphi) \sd x = \int_V e(\ve,\varphi,\nabla \varphi) \sd x,
\end{equation*}
where $e(\ve,\varphi,\nabla \varphi):= \hrho(\varphi) \frac{|\ve|^2}2+ f(\varphi,\nabla \varphi)$.

\subsection[Local Dissipation Inequality]{Derivation based on a Local Dissipation Inequality and Microstresses}

In the following $V(t)\subseteq \Omega$ shall denote an arbitrary volume that is transported with the flow,  i.e., the exterior normal velocity of $V(t)$ is given by $\nl\cdot \ve(t)$ on $\partial V(t)$.

In order to describe the change of the free energy due to diffusion, we introduce a
\emph{chemical potential} $\mu$ and the outer unit normal $\bfnu$
to $\partial V(t)$ such that
\begin{equation*}
     -\int_{\partial V(t)}  \mu \, \vc{J}\cdot \bfnu \dsm - \int_{\partial V(t)}  \tfrac{|\ve|^2}2 \tilde{\vc{J}}\cdot \nl \dsm 
\end{equation*}
is the energy transported into $V(t)$ by diffusion and $\dsm$ denotes integration with respect to the 
surface measure. Here the term 
 $-\int_{\partial V(t)}  \mu \, \vc{J}\cdot \bfnu \dsm$
describes the change of free energy. The kinetic energy is transported
by the velocity $\ve+ \tilde{\vc{J}}/\rho$ and 
the term $- \int_{\partial V(t)}  \tfrac{|\ve|^2}2 \tilde{\vc{J}}\cdot \nl \dsm$ 
describes the change of kinetic energy due to diffusion,
see also \cite{AltEntropy}, Section 10.

\noindent
{\it Surface forces:} Moreover, we assume the existence of a generalized (vectorial) surface force $\boldsymbol{\xi}$ such that 
\begin{equation*}
  \int_{\partial V(t)} \dot{\varphi}\, \boldsymbol{\xi} \cdot \nl \dsm
\end{equation*}
represents the working due to microscopic stresses. 
Above and in the following $\dot{\varphi}$ is the material derivative
$\partial_t \varphi +\ve\cdot \nabla \varphi$. Finally, we note that 
\begin{equation*}
  \int_{\partial V(t)} (\tn{T} \nl) \cdot \ve \dsm
\end{equation*}
describes the working in a given volume $V(t)$  due to the macroscopic stresses in the fluid.\\[1ex]
\noindent
{\it Second law of thermodynamics/local dissipation inequality:}
Similar as in \cite{GurtinTwoPhase}, we assume the following dissipation inequality, which is the appropriate formulation of the second law of thermodynamics in an isothermal situation:
\begin{eqnarray*}
  \lefteqn{\frac{d}{dt} \int_{V(t)} e(\ve,\varphi,\nabla \varphi) \sd x}\\
 &\leq& \int_{\partial V(t)} (\tn{T} \nl)\cdot \ve \dsm +  \int_{\partial V(t)} \dot{\varphi}\, \boldsymbol\xi \cdot \nl \dsm - \int_{\partial V(t)}  \mu \vc{J}\cdot \nl \dsm - \int_{\partial V(t)}  \tfrac{|\ve|^2}2 \tilde{\vc{J}}\cdot \nl \dsm
\end{eqnarray*}
for every volume $V(t)$ transported with the flow. This means that the change of total energy in time is bounded by the working due to macroscopic and microscopic stresses and the change of energy due to diffusion.

We recall the transport theorem, see e.g. Liu~\cite[Theorem~2.1.]{Liu}:
\begin{equation}\label{eq:transport}
  \frac{d}{dt} \int_{V(t)} f \sd x = \int_{V(t)} \partial_t f \sd x + \int_{\partial V(t)}\nl \cdot \ve f \dsm
 = \int_{V(t)} \left(\partial_t f+ \Div (\ve f)\right) \sd x 
\end{equation}
where $V(t)$ is transported with the flow described by $\ve$.
 Therefore the equivalent local form is 
\begin{equation}\label{eq:LocalDiss}
  \partial_t e + \ve\cdot \nabla e - \Div (\tn{T}^T \ve)+ \Div \left(\tfrac{|\ve|^2}2 \tilde{\vc{J}}\right) - \Div (\dot{\varphi} \boldsymbol{\xi})+ \Div (\mu \vc{J})=:-\mathcal{D} \leq 0.
\end{equation}
Using 
(\ref{eq:DivFree'}),
(\ref{eq:DiffEq1}) and (\ref{eq:ConservationLinMomentum}), we will simplify $\mathcal{D}$. First of all, multiplying (\ref{eq:ConservationLinMomentum}) with $\vc{v}$ and using (\ref{eq:DiffEq1'}), we obtain
\begin{eqnarray}\nonumber
\partial_t\left(\rho\tfrac{|\bu|^2}{2})+\bu\cdot\nabla(\rho\tfrac{|\bu|^2}{2}\right)
&=&
(\partial_t\rho+\bu\cdot\nabla\rho)\tfrac{|\bu|^2}{2}+\rho(\partial_t\bu+(\bu\cdot\nabla)\bu)\cdot\bu\\\nonumber
&=&
-(\Div\tilde{\vc{J}})\tfrac{|\bu|^2}{2}+(\Div\widetilde{\tn{S}}-\nabla
p-(\tilde{\vc{J}}\cdot\nabla)\bu)\cdot\bu\\\label{eq:Id1}
&=&
\Div\left(-\tfrac{1}{2}|\bu|^2\tilde{\vc{J}}+\widetilde{\tn{S}}^T\bu -p
\bu\right)-\widetilde{\tn{S}}:\nabla\bu\,.
\end{eqnarray}
Moreover, 
\begin{equation}\label{eq:Id2}
    \partial_t f +\vc{v}\cdot \nabla f =
    \frac{\partial f}{\partial \varphi} \dot{\varphi} + \frac{\partial f}{\partial \nabla \varphi} \cdot (\nabla \varphi)^{\cdot} 
\end{equation}
and
\begin{eqnarray*}
  \Div (\dot{\varphi} \boldsymbol{\xi})- \Div (\mu \vc{J}) &=& \dot{\varphi} \Div \boldsymbol{\xi} + \nabla (\dot{\varphi})\cdot \boldsymbol{\xi} + \mu \frac{\partial \widehat{\rho c}}{\partial \varphi}\dot{\varphi} - \nabla \mu \cdot \vc{J} \\ 
&=& (\tfrac{\partial \widehat{\rho c}}{\partial \varphi }\mu+ \Div \boldsymbol{\xi}) \dot{\varphi} + (\nabla \varphi)^{\cdot}\cdot \boldsymbol{\xi} + \nabla \ve: (\nabla \varphi \otimes \boldsymbol{\xi})  - \nabla \mu \cdot \vc{J},
\end{eqnarray*}
because of (\ref{eq:DiffEq1'}) and where we have used 
\begin{equation}\label{eq:Id3}
  \partial_{x_j} \dot{\varphi}= \partial_t \partial_{x_j} \varphi + \ve\cdot \nabla \partial_{x_j}\varphi + \partial_{x_j} \ve\cdot \nabla \varphi= (\partial_{x_j}\varphi)^{\cdot} + \partial_{x_j} \ve\cdot \nabla \varphi.
\end{equation}
Thus we conclude that (\ref{eq:LocalDiss}) is equivalent to
\begin{eqnarray}\nonumber
  \lefteqn{-\mathcal{D}=\left(\frac{\partial f}{\partial \varphi}- \Div \boldsymbol{\xi}-\frac{\partial \widehat{\rho c}}{\partial \varphi}\mu\right)\dot{\varphi}}\\\label{eq:DissInequality}
&& + \left(\frac{\partial f}{\partial \nabla \varphi}-\boldsymbol{\xi} \right)\cdot(\nabla \varphi)^{\cdot}  - (\widetilde{\tn{S}}+ \nabla \varphi\otimes \boldsymbol{\xi}):\nabla \ve + \nabla \mu \cdot \vc{J} \leq 0,
\end{eqnarray}
where we have used $\Div \ve=0$.

In order to motivate the constitutive assumptions, we will derive some
restrictions for the constitutive relations specifying
$\widetilde{\tn{S}},\mu, \vc{J}, \xi$ by an argument typical for rational
continuum mechanics: To this end, we assume that
$\widetilde{\tn{S}},\vc{J},\boldsymbol{\xi}$ are functions of $D\ve,
\varphi,\nabla \varphi, \mu, \nabla \mu$ only. Moreover, we assume that
$\widetilde{\tn{S}}$ is symmetric.
 Invoking general external forces and mass supplies in the equations, one argues that
$\varphi,\dot{\varphi}, \nabla  \varphi, (\nabla \varphi)^{\cdot},\mu,$
$ \nabla \mu, (\nabla \ve- \frac13 \Div \ve \tn{I})$ can attain arbitrary values for a given point in space and time and since $f$ and $\boldsymbol{\xi}$ do not depend on $\dot{\varphi}, (\nabla \varphi)^{\cdot}$, we conclude
from (\ref{eq:DissInequality}) that
\begin{equation*}
  \boldsymbol{\xi} -  \frac{\partial f}{\partial \nabla \varphi}(\varphi,\nabla \varphi) =0 
\end{equation*}
necessarily. In particular, $\boldsymbol{\xi}$ depends only on $\varphi,\nabla \varphi$.
Hence (\ref{eq:DissInequality}) reduces to
\begin{eqnarray}\nonumber
  \lefteqn{-\mathcal{D}=\left(\frac{\partial f}{\partial \varphi}- \Div \boldsymbol{\xi}-\frac{\partial \widehat{\rho c}}{\partial \varphi}\mu\right)\dot{\varphi}}\\\label{eq:DissInequality2}
&&-  (\widetilde{\tn{S}}+\nabla \varphi\otimes \boldsymbol{\xi}): D\ve  -(\nabla\varphi \otimes \boldsymbol{\xi}): \frac12(\nabla \ve-\nabla \ve^T) + \nabla \mu \cdot \vc{J}\leq 0,
\end{eqnarray}
where $D\ve= \frac12 (\nabla \ve+ \nabla \ve^T)$.
Since the skew part of $\nabla \ve$ can attain arbitrary values independent of $D\ve$ and 
\begin{equation*}
(\nabla\varphi \otimes \boldsymbol{\xi}): \frac12(\nabla \ve-\nabla \ve^T)= \frac12\left( \nabla\varphi \otimes \boldsymbol{\xi}- \boldsymbol{\xi}\otimes \nabla\varphi \right): \frac12(\nabla \ve-\nabla \ve^T),  
\end{equation*}
 we conclude that $\tfrac12\left( \nabla\varphi \otimes \boldsymbol{\xi}- \boldsymbol{\xi}\otimes \nabla\varphi \right)=0$ and therefore
\begin{equation*}
  \boldsymbol{\xi}\otimes \nabla \varphi = \nabla \varphi \otimes \boldsymbol{\xi}.
\end{equation*}
Thus $|\boldsymbol{\xi}|^2 |\nabla \varphi|^2 = (\nabla \varphi\cdot \boldsymbol{\xi})^2$ and therefore 
\begin{equation}\label{eq:rel1}
 \boldsymbol{\xi}(\varphi,\nabla \varphi)=a(\varphi,\nabla \varphi) \nabla \varphi = \frac{\partial f}{\partial \nabla \varphi}(\varphi,\nabla \varphi) 
\end{equation}
for some $a(\varphi,\nabla \varphi)$.

On the other hand,
 the first term after the equality sign in
 (\ref{eq:DissInequality2}) is linear in $\dot{\varphi}$ and  $f,\mathbf J$ and $\mu$ are assumed to be independent of $\dot{\varphi}$. Therefore the first term in (\ref{eq:DissInequality2}) has to vanish to satisfy (\ref{eq:DissInequality2}) in general and
\begin{equation*}
 \frac{\partial \hat{\rho c}}{\partial \varphi}\mu = \frac{\partial f}{\partial \varphi} -\Div (a(\varphi,\nabla \varphi)\nabla \varphi).
\end{equation*}
Finally, the local dissipation inequality is satisfied if and only if
\begin{equation}\label{eq:DissInequal2}
 \mathcal{D}=  \left(\widetilde{\tn{S}}+ \nabla \varphi \otimes \frac{\partial f}{\partial \nabla \varphi}\right):D \ve - \nabla \mu \cdot \vc{J} \geq 0.
\end{equation}
Here $\widetilde{\tn{S}}+  \nabla \varphi \otimes \frac{\partial f}{\partial \nabla \varphi} $ is also called viscous stress tensor since it corresponds to irreversible changes of the energy due to friction in the fluids.

\noindent
{\it Constitutive assumptions:} Motivated by Newton's rheological law, cf. e.g. \cite[Section~4.2.2]{Liu}, we assume that 
\begin{equation*}
  \widetilde{\tn{S}}+  \nabla \varphi\otimes \frac{\partial f}{\partial \nabla \varphi}  = 2\eta(\varphi) D\ve 
\end{equation*}
for some function $\eta(\varphi)\geq 0$.

Finally, we choose 
$\mathbf J (\varphi,\nabla \varphi,\mu, \nabla\mu)$ in the form 
\begin{equation*}
  \mathbf J(\varphi,\nabla \varphi,\mu, \nabla\mu)= -\tilde{m}(\varphi)\nabla \mu, 
\end{equation*} 
where $\tilde{m}(\varphi) \geq 0$, which corresponds to a generalized Fick's law. We remark that $\mathbf J$ can  be chosen to be nonlinear with respect to $\nabla \mu$ as long as (\ref{eq:DissInequal2}) is fulfilled.

Summing up, we derived the following diffuse interface model:
\begin{alignat}{2}\label{eq:FirstLT1}
   \rho\partial_t \ve + ((\rho \ve+\tilde{\vc{J}})\cdot \nabla) \ve -\Div \left( 2\eta(\varphi) D\ve\right)+ \nabla p&= -\Div (a(\varphi,\nabla \varphi) \nabla \varphi \otimes \nabla \varphi),\\ 
\label{eq:FirstLT2}
  \Div \ve&= 0, \\
\label{eq:FirstLT3}
  \partial_t \widehat{\rho c}(\varphi)+ \ve\cdot \nabla \widehat{\rho c}(\varphi)  &= \Div(\tilde{m}(\varphi) \nabla \mu)
\end{alignat}
together with
\begin{equation}\label{eq:FirstLT4}
 \frac{\partial \widehat{\rho c}}{\partial \varphi}\mu = \frac{\partial f}{\partial \varphi}(\varphi,\nabla \varphi) -\Div (a(\varphi,\nabla \varphi) \nabla \varphi),
\end{equation}
where $a(\varphi,\nabla \varphi)$ satisfies (\ref{eq:rel1}) and $\tilde{\vc{J}}= -\frac{\tilde{\rho}_2-\tilde{\rho}_1}{\tilde{\rho}_2+\tilde{\rho}_1} \tilde{m}(\varphi)\nabla \mu $  due to \eqref{eq:RelationJ}. 
Assuming that the normal component of $\ve$ and the normal derivative of $\mu$ vanishes on the boundary of the fluid domain $\Om$, 
$\int_\Om\widehat{\rho c}(\cdot,t)\sd x$ is a constant in time. By \eqref{eq:neu}
and \eqref{eq:neu2}, we may express both mass densities as affine linear functions of $\widehat{\rho c}$. Hence, the total mass 
$\int_\Omega \rho_i \, dx$, $i=1,2$, of each liquid component is conserved. 

If $\varphi = c$ is the mass concentration difference, then $\widehat{\rho c}(c)=\hrho(c)c$, and we obtain
\begin{alignat}{2}\label{eq:FirstLT1b}
\rho\partial_t \ve + ((\rho \ve+\tilde{\vc{J}})\cdot \nabla)\ve -\Div \left(
2\eta(c) D\ve\right)+ \nabla p&= -\Div (a(c,\nabla c) \nabla c
\otimes \nabla c),\\ 
\label{eq:FirstLT2b}
  \Div \ve&= 0, \\
\label{eq:FirstLT3b}
  \partial_t (\rho c)+ \ve\cdot \nabla 
(\rho c)  &= \Div(\tilde{m}(c) \nabla \mu)
\end{alignat}
together with
\begin{equation}
 \frac{\partial (\widehat{\rho c})}{\partial c}\mu = \frac{\partial f}{\partial c}(c,\nabla c) -\Div (a(c,\nabla c) \nabla c)
\end{equation}
and $\tilde{\vc{J}}= -\frac{\tilde{\rho}_2-\tilde{\rho}_1}{\tilde{\rho}_2+\tilde{\rho}_1} \tilde{m}(c)\nabla \mu $.
Here
\begin{equation*}
  \frac{\partial (\widehat{\rho c})}{\partial c}= \alpha\rho^2 \quad \text{with}\ \alpha = \frac1{2\tilde{\rho_1}} +\frac1{2\tilde{\rho_2}}. 
\end{equation*}

In the case that $\varphi=\rho c$ is the density difference, we have $\widehat{\rho c}(\varphi)=\varphi$, $\frac{\partial \widehat{\rho c}}{\partial \varphi}=1$, and therefore 
\begin{alignat*}{2}
  \rho\partial_t \ve + ((\rho \ve+\tilde{\vc{J}})\cdot \nabla)\ve  -\Div
 \left( 2\eta(\varphi) D\ve\right)
+ \nabla p&= -\Div (a(\varphi,\nabla \varphi) \nabla
\varphi \otimes \nabla \varphi),\\ 
  \Div \ve&= 0, \\
  \partial_t \varphi+ \ve\cdot \nabla \varphi  &= \Div(\tilde{m}(\varphi) \nabla \mu)
\end{alignat*}
and
\begin{equation*}
 \mu = \frac{\partial f}{\partial \varphi}(\varphi,\nabla \varphi) -\Div (a(\varphi,\nabla \varphi) \nabla \varphi).
\end{equation*}

Finally, in the case that $\varphi=u_2-u_1$ is the difference of volume fractions, we have $\widehat{\rho
c}(\varphi)=\frac{\tilde\rho_2-\tilde\rho_1}{2}+\frac{\tilde\rho_1+\tilde\rho_2}{2}\varphi$ and we obtain the
system
\begin{alignat}{2}\label{eq:FirstLT1c}
  \rho\partial_t \ve + ((\rho \ve+\tilde{\vc{J}})\cdot \nabla)\ve  -\Div
 \left( 2\eta(\varphi) D\ve\right)
+ \nabla p&= -\Div (a(\varphi,\nabla \varphi) \nabla
\varphi \otimes \nabla \varphi),\\ 
\label{eq:FirstLT2c}
  \Div \ve&= 0, \\
\label{eq:FirstLT3c}
 \partial_t \varphi+ \ve\cdot \nabla \varphi  &= -\Div \J, 
\end{alignat}
together with
\begin{alignat}{2}\label{eq:FirstLT4c}
 \mu_\varphi &=  \frac{\partial f}{\partial \varphi}(\varphi,\nabla \varphi) -\Div (a(\varphi,\nabla \varphi)
\nabla \varphi) 
\end{alignat}
where we use a rescaled flux
$\J = \left(\frac{\tilde\rho_1+\tilde\rho_2}{2}\right)^{-1} {\bf J}$,
and a rescaled chemical potential
$\mu_\varphi= \frac{\tilde\rho_1+\tilde\rho_2}{2} \mu$. We hence obtain
\begin{equation*}
\J =-m(\varphi)
 \nabla \mu_\varphi,\quad \tilde{\vc{J}} =- \tfrac{\tilde{\rho}_2-\tilde{\rho}_1}2 m(\varphi)\nabla \mu_\varphi  
\end{equation*}
where $m(\varphi)=\left(\frac{\tilde\rho_1+\tilde\rho_2}{2}\right)^{-2}\tilde{m}(\varphi)$.
Usually we take the difference of volume fractions as order parameter. This has the advantage that
the mass difference depends linearly on the order parameter and as usual in phase field
models the values $\varphi= \pm 1$ correspond to unmixed ``pure'' phases.

Finally we remark that (\ref{eq:FirstLT3c}) implies the mass balance
$$\partial_t \rho + \Div  (\rho \ve + \tilde{ \vc{J}} ) =0. $$
Using this we can rewrite the momentum balance (\ref{eq:FirstLT1c}) as
\begin{equation*}
 \partial_t(\rho \ve) + \Div (\rho \ve\otimes \ve) + \Div (\ve\otimes
\tilde{ \vc{J}  }) -\Div
(2\eta(\varphi)D\ve)+ \nabla p
= -\Div ( a(\varphi,\nabla \varphi) \nabla
\varphi \otimes \nabla \varphi) \, .  
\end{equation*}

\begin{rem}
  For the previous derivation it is essential to use the frame invariant
  version of the momentum equation \eqref{mom1} in order to derive a frame
  invariant model. If one does the derivation in the same manner using
  \eqref{eq:ConservationLinMomentum} instead, the resulting model is 
\begin{alignat}{2}
\label{non1}
   \partial_t(\rho \ve) + \Div (\rho \ve\otimes \ve) -\Div \left( 2\eta(\varphi) D\ve\right)+ \nabla p&= -\Div (a(\varphi,\nabla \varphi) \nabla \varphi \otimes \nabla \varphi),\\ 
  \Div \ve&= 0, \\
  \partial_t \widehat{\rho c}(\varphi)+ \ve\cdot \nabla \widehat{\rho c}(\varphi)  &= \Div(\tilde{m}(\varphi) \nabla \mu)
\end{alignat}
together with
\begin{equation}
\label{non4}
 \frac{\partial \widehat{\rho c}}{\partial \varphi}\mu = \frac{\partial f}{\partial \varphi}(\varphi,\nabla \varphi) -\Div (a(\varphi,\nabla \varphi) \nabla \varphi) - \frac{\partial \rho}{\partial \varphi} \frac{|\ve|^2}2,
\end{equation}
cf. \cite{ModifiedModelH1} for details. This model is not frame invariant since 
 the chemical potential contains with the term  $\frac{\partial \rho}{\partial
   \varphi} \frac{|\ve|^2}2$ a non-objective scalar. 
\end{rem}

\subsection[Local Dissipation Inequality]{Derivation based on a Local Dissipation Inequality and the Lagrange Multiplier Approach}

It is also possible to exploit a dissipation inequality without introducing generalized surface forces.
We now assume a dissipation inequality
\begin{equation*}
  \frac{d}{dt} \int_{V(t)} e(\ve,\varphi,\nabla \varphi) \sd x + \int_{\partial V(t)} \vc{J}_e\cdot \boldsymbol{\nu}\, \dsm
\leq 0
\end{equation*}
for every volume $V(t)$ transported with the flow for some general \emph{energy flux} $\vc{J}_e$, which will be specified later.

 Then the equivalent local form is 
\begin{equation}\label{eq:LocalDissB}
  \partial_t e + \ve\cdot \nabla e + \Div \vc{J}_e\leq 0. 
\end{equation}
Because of the conservation law
(\ref{eq:Flux}), we conclude that for every scalar function $\lambda_\varphi$ the inequality
\begin{eqnarray}
  -\mathcal{D}:=\partial_t e + \ve\cdot \nabla e + \Div \vc{J}_e \label{eq:LocalDiss2}
- \lambda_\varphi \left(\partial_t(\rho c )+ \ve\cdot\nabla (\rho c) +\Div \vc{J}\right)&\leq& 0 
\end{eqnarray}
has to be valid.

Using  (\ref{eq:ConservationLinMomentum}), (\ref{eq:Id1}), (\ref{eq:Id2}) we obtain 
that (\ref{eq:LocalDiss2}) is equivalent to
\begin{eqnarray}\nonumber
  \lefteqn{-\mathcal{D}=\left(\frac{\partial f}{\partial \varphi}- \lambda_\varphi\frac{\partial (\rho c)}{\partial \varphi}\right)\dot{\varphi}}\\\nonumber
&& + \left(\frac{\partial f}{\partial \nabla \varphi} \right)\cdot (\nabla \varphi)^{\cdot}+\Div \left(-\tfrac{|\ve|^2}2 \tilde{\vc{J}}+\tn{T}\cdot \ve-\lambda_\varphi \vc{J}+ \vc{J}_e\right)  - \widetilde{\tn{S}}:\nabla \ve + \nabla \lambda_\varphi \cdot \vc{J} \leq 0,
\end{eqnarray}
where we have used $\Div \ve=0$. 
Making use of (\ref{eq:Id3})
we conclude that the latter inequality is equivalent to
\begin{eqnarray}\nonumber
  -\mathcal{D}&=&\left(\frac{\partial f}{\partial \varphi}- \lambda_\varphi\frac{\partial (\rho c)}{\partial \varphi}- \Div \left(\frac{\partial f}{\partial \nabla \varphi}\right)\right)\dot{\varphi}- \left(\widetilde{\tn{S}}+\nabla \varphi\otimes \frac{\partial f}{\partial \nabla \varphi}\right):\nabla \ve\\\label{eq:DissInequalityB}
&& +\Div \left(-\tfrac{|\ve|^2}2 \tilde{\vc{J}}+\tn{T}\cdot \ve-\lambda_\varphi \vc{J}+\frac{\partial f}{\partial\nabla \varphi}\dot \varphi+ \vc{J}_e\right)   + \nabla \lambda_\varphi \cdot \vc{J} \leq 0,
\end{eqnarray}
where  again the equation $\Div \ve=0$ was used.

If we assume now that $\widetilde{\tn{S}},\lambda_\varphi$, and $-\tfrac{|\ve|^2}2 \tilde{\vc{J}}+ \tn{T}\cdot \ve -\lambda_\varphi \vc{J}
+\frac{\partial f}{\partial\nabla \varphi}\dot \varphi-\vc{J}_e $ are independent of $\dot{\varphi}$, we
conclude that the first term after the equality sign in (\ref{eq:DissInequalityB}) has to vanish for all values of $\dot{\varphi}$. Thus
\begin{equation*}
  \lambda_\varphi\frac{\partial (\rho c)}{\partial \varphi}= \frac{\partial f}{\partial \varphi}- \Div \left(\frac{\partial f}{\partial \nabla \varphi}\right).
\end{equation*}
If we denote $\lambda_\varphi=\mu$, then we obtain the same identity for the ``chemical potential'' $\mu$ as before. Moreover, if we now specify the energy flux as
\begin{equation*}
  \vc{J}_e=\tfrac{|\ve|^2}2 \tilde{\vc{J}}-\tn{T}\cdot \ve+\mu \vc{J} - \frac{\partial f}{\partial \nabla \varphi}\dot{\varphi},
\end{equation*}
we end up with the local dissipation inequality
\begin{equation*}
 \mathcal{D}=  \left(\widetilde{\tn{S}}+ \nabla \varphi \otimes \frac{\partial f}{\partial \nabla \varphi}\right):D \ve - \nabla \mu \cdot \vc{J} \geq 0,
\end{equation*}
which is just (\ref{eq:DissInequal2}). Thus we can derive the 
 model (\ref{eq:FirstLT1})-(\ref{eq:FirstLT4}) as before by making the same constitutive assumptions.
\section{Onsager's Variational Principle -- a Third Approach to Derive
  Thermodynamically Consistent Models} \label{sec:Onsager}

In this section, we follow a third pathway to obtain diffuse interface
models in the spirit of (\ref{eq:FirstLT1c})-(\ref{eq:FirstLT3c}) and in
agreement with the postulations of thermodynamics. It is based on
Onsager's variational principle, see
\cite{ons31} and  \cite{ecketal}, \cite{fogrjo}, \cite{qws} for applications
in multi-phase flow.
To widen the range of applications,  we discuss two additional features. We
allow for gravitational forces or, alternatively, we include
the transport of a soluble species across fluidic interfaces as an additional
effect.  For simplicity,
we refrain ourselves in the second case to a species  that does not
influence the surface tension at the interface.
Adopting the notation from the previous sections, the order parameter will be
denoted by  $\varphi$
and we take it to be the difference $u_2-u_1$ of the volume fractions $u_j$,
$j=1,2$, of the two liquids
involved. Hence,
\begin{equation}
\label{rho}
\rho(\varphi)=\frac{\tilde{\rho}_2-\tilde{\rho}_1}2\varphi+\frac{\tilde{\rho}_1+\tilde{\rho}_2}2
\end{equation}
where $\tilde{\rho}_1$ and $\tilde{\rho}_2$ are the specific densities of
liquid $1$ and $2$,
respectively.
By $w$, we denote the mass density of the soluble species which we assume to
be dilute.

Our starting point is the following set of general evolution equations,
compare
\eqref{gg-new} and \eqref{eq:momentum2},
\begin{eqnarray}\label{equa}
  \rho\partial_t\bu+\left(\left(\rho\bu+\dichtabl\J\right)\cdot\nabla\right)\bu
-\Div \, \stress+\nabla p&=&\K,\\
  \Div\,\bu&=&0,\\
  \partial_t\varphi+\bu\cdot\nabla\varphi +\Div\, \J &=&0,\\
\partial_t w+\bu\cdot\nabla w +\Div\, \Jw &=&0 \label{consw}
\end{eqnarray}
with $\rho(\cdot)$ as in \eqref{rho} and $ \J=
\left(\frac{\tilde\rho_1+\tilde\rho_2}{2}\right)^{-1}\mathbf J  $
a rescaled flux, compare the discussion after (\ref{eq:FirstLT4c}).
 The stress tensor $ \stress$ is symmetric,
and $\K$ denotes the force density.
The additional equation (\ref{consw}) is the mass balance of the soluble
species and
$\Jw$ is the corresponding mass flux.
 These equations are supposed to hold in a space-time cylinder
$\Omega\times(0,T)$ with $\Omega\subset\R^d$ being the domain where the
process takes place. Conservation of mass requires that the normal components
of $\J$ and of $\Jw$ vanish on $\partial\Omega.$
As  free energy, we choose
$$
\int_\Om f(\varphi,\nabla \varphi) \sd x+\int_\Om \{g(w)+ \beta(\varphi)w\}\sd
x
$$
with
$f(z,p)=f_1(z)+f_2(p).$
Here, $g(\cdot)$ is an entropic term, and $\beta(\varphi)$ attains for
$\varphi\leq-1$ or $\varphi\geq 1$ the
values $\beta_1$ or $\beta_2$, respectively. In a sharp-interface limit, these
parameters will reappear through the Henry jump condition
\begin{equation}
\label{henry}
\frac{w_1}{w_2}=\EXP(\beta_2-\beta_1)
\end{equation}
 at the interface separating the two phases
(cf. Subsection \ref{leadorder}).
The total energy is given by the sum of kinetic and free energy, hence
$$
F=\int_\Omega \frac{\rho(\varphi)}{2} |\bu|^2\sd x+\int_\Om f(\varphi,\nabla
\varphi)\sd x
+\int_\Om \{g(w)+\beta(\varphi)w\}\sd x
.
$$
The time derivative of the free energy is given as
\begin{equation*}\begin{split}
\frac{d F}{d
  t}=&\frac12\int_\Om\frac{\partial\rho}{\partial\varphi}|\bu|^2\partial_t\varphi\sd
x+\int_\Om\frac{\partial
  f}{\partial \varphi} \partial_t\varphi\sd x-\int_\Om
\Div\left(\frac{\partial
    f}{\partial\nabla
    \varphi}\right)\partial_t\varphi\sd x \\ 
& +\int_\Om\rho(\varphi)\bu\cdot\partial_t\bu\sd x
+\int_\Om(g'(w)+\beta(\varphi))\partial_tw\sd x 
  +\int_\Om\beta'(\varphi)w\partial_t\varphi\sd x, \end{split}
\end{equation*}
where we assumed that the normal part of $\frac{\partial
  f}{\partial \nabla\varphi}$ vanishes on $\partial \Omega$.
Observe that
$$
\partial_t\bu=\frac1\rho\left\{\K+\Div\,\stress-\left(\left(\rho\bu+\dichtabl\J\right)\cdot\nabla\right)\bu-\nabla
p \right\}.
$$
Inserting the evolution equations and integrating by parts gives, assuming
$\ve\equiv0$ on $\partial \Omega$,
\begin{equation*}\begin{split}
\frac{d}{dt}F=&-\frac12\int_\Om\dichtabl|\bu|^2(\bu\cdot\nabla
\varphi+\Div\,\J
)\sd x-\int_\Om\frac{\partial f}{\partial \varphi}(\bu\cdot\nabla
\varphi+\Div\,\J )\sd x\\
&+\int_\Om \Div\left(\frac{\partial f}{\partial \nabla
    \varphi}\right)(\bu\cdot\nabla
\varphi+\Div\,\J )\sd x-\int_\Om\beta'(\varphi)w\left(\bu\cdot\nabla\varphi
  +\Div\J\right)\sd x\\
&+\int_\Om
 \bu\cdot \left\{\K+ \Div \,
  \stress-\left(\left(\rho\bu+\dichtabl\J\right)\cdot\nabla\right)\bu
-\nabla p\right\}\sd x\\
&-\int_\Om\left(g'(w)+\beta(\varphi)\right)\left(\bu\cdot\nabla
w+\Div\Jw\right)\sd x\\
=&\int_\Om\bu\cdot\nabla \varphi\left\{-\frac{\partial f}{\partial
    \varphi}+\Div\left(\frac{\partial f}{\partial\nabla
      \varphi}\right)-\beta'(\varphi) w\right\}\sd x\\
&+\int_\Om\bu\cdot\K\sd x-\int_\Om\bu\cdot\nabla
w\left(g'(w)+\beta(\varphi)\right)\sd x\\
&+\int_\Om\J\cdot\nabla\mu_\varphi\sd x-\int_\Om D\bu:\stress\sd
x+\int_\Om\Jw\cdot\nabla\mu_w \sd x.
\end{split}
\end{equation*}
Here, we used
$$
\int_\Om\bu\cdot\left(\left(\rho\bu+\dichtabl\J\right)\cdot\nabla\right)\bu\sd x=-\frac12\int_\Om\dichtabl|\bu|^2\left(\Div\J+\nabla\varphi\cdot\bu\right)\sd x,
$$
which -- as a consequence of the no-flux boundary conditions -- follows easily by
integration by parts.
Moreover, we abbreviated
$$\mu_\varphi:=-\Div\left(\frac{\partial f}{\partial \nabla
\varphi}\right)+\frac{\partial
  f}{\partial \varphi} +\beta'(\varphi)w$$
and
$$\mu_w:=g'(w)+\beta(\varphi)$$
to denote the chemical potentials corresponding to $\varphi$ and $w,$
respectively.
Recall that in general entropy production is due to external force fields and
to gradients of velocity  and
of  chemical potentials, see
\cite{deGroot}, Chapter 3. However, if the specific densities of the external
forces\footnote{i.e. the quotient of force density and mass density.} acting on
the different species are identical, then those forces do not contribute to
entropy production. Therefore, if no other external forces than gravity forces
are applied, we may identify the rate of change of the mechanical work with
\begin{equation}\label{mechanics}
\frac{dW}{dt}= -\int_\Om\bu\cdot \nabla \varphi\mu_\varphi\sd
x-\int_\Om\bu\cdot \nabla w
\mu_w\sd x+\int_\Om\bu\cdot\K\sd x\sd x.
\end{equation}
As $$\frac{dW}{dt}=\begin{cases}\int_\Om \mathbf K_{\mathrm{grav}}\cdot\bu\sd
x & \text{case with gravitational forces, }\\
0 & \text{case with soluble species,}
\end{cases}$$
equation \eqref{mechanics} is satisfied if
$$\K=\begin{cases}
\mu_\varphi\nabla
\varphi+{\mathbf K_{\mathrm{grav}}} &\text{case with gravitational
forces},\\
\mu_\varphi\nabla
\varphi+\mu_w\nabla w & \text{case with soluble species}.
\end{cases}$$
Here, $\mathbf K_{\mathrm{grav}}$ denotes the gravitational force.
\newline
To determine the fluxes $\J, \Jw$ and the stress tensor $\stress, $ we
introduce the
dissipation functional
\begin{equation}
\label{diss-func}
\Phi (\J ,\Jw,\stress):=\int_\Omega
\left\{\frac{|\J
|^2}{2M(\varphi)}+\frac{|\stress|^2}{4\eta(\varphi)}+\frac{|\Jw|^2}{2K(\varphi)w}\right\}
\sd x.
\end{equation}
We use Onsager's variational principle which postulates
$$\delta_{(\J,\Jw,\stress)}\left(\frac{dF}{dt}(\J,\Jw
,\stress)+\Phi(\J,\Jw,\stress)\right)\stackrel{!}{=}0.$$
This gives $\J =-M(\varphi)\nabla\mu_\varphi$,
$\Jw=-K(\varphi)w\nabla\mu_w,$ and
$\stress=2\eta(\varphi)\frac{\nabla\bu+\nabla\bu^T}2.$
Altogether,  when a soluble species is around, we end up with
\begin{eqnarray}
\nonumber
\rho\partial_t \bu+\left(\left(\rho\bu+\dichtabl\J\right)\cdot\nabla\right)\bu
 \\\label{sol1}
-\Div\big(2\eta(\varphi)D\bu\big)+\nabla
p&=&\mu_\varphi\nabla\varphi+\mu_w\nabla w,\\ \label{sol2}
\Div \bu &=& 0, \\\label{sol3}
\partial_t \varphi +
\bu\cdot\nabla\varphi-\Div\big(M(\varphi)\nabla\mu_\varphi\big)&=&0,
\\\label{sol4}
\partial_t w+\bu\cdot \nabla w-\Div\left(K(\varphi)w\nabla\mu_w\right) &=& 0,
\\\label{sol5}
\mu_\varphi=-\Div\left(\frac{\partial f}{\partial\nabla
    \varphi}\right)+\frac{\partial f}{\partial \varphi}&+&\beta'(\varphi)w, \\
\label{sol6}
\mu_w&=&g'(w)+\beta(\varphi)
 \end{eqnarray}
where $D \bu=\frac12\big(\nabla\bu+\nabla\bu^T\big)$ and
$\rho(\varphi)=\frac{\tilde{\rho}_2-\tilde{\rho}_1}2\varphi+\frac{\tilde{\rho}_1+\tilde{\rho}_2}2$.
The evolution equation for $w$ becomes
\begin{equation}
\label{w-explizit}
  \partial_t w+\bu \cdot\nabla w =
\Div(K(\varphi)w\nabla(g'(w)+\beta(\varphi))).
\end{equation}
In the case of gravitational forces, the system has to be changed accordingly.

\vspace{ 2mm}
\noindent
{\it Remarks:}
\begin{itemize}
\item If we choose the viscosity parameter in \eqref{diss-func} as
$\eta(\varphi, D\bu)$,
  non-newtonian effects, e.g. shear thinning or shear thickening, can be
  included as 
  well. 
\item If we choose $g(w)=w(\log w -1)$, the equations
\eqref{sol4},\eqref{sol6} result in the diffusion equation
  \begin{equation*}
    \partial_t w+\bu\cdot \nabla w = \Div (K(\varphi)(\nabla w+w\nabla
\beta(\varphi))).
  \end{equation*}
\item
The interfacial force
  term $\vc{K}= \mu_\varphi \nabla\varphi+ \mu_w \nabla w$ can equivalently be
written
  as $$
\nabla (f(\varphi,\nabla\varphi) + g(w) +\beta(\varphi)
w)-\Div\left(\nabla\varphi\otimes\frac{\partial
  f(\varphi,\nabla\varphi)}{\partial\nabla\varphi}\right).$$
Here, the first term is of pressure type whereas $\nabla\varphi\otimes
  \frac{\partial f}{\partial\nabla\varphi} $ provides an additional stress
  tensor contribution representing interfacial forces. We can hence conclude
that the
derivations in Sections \ref{eq:Derivation} and \ref{sec:Onsager}
up to a reinterpretation of the pressure lead to the same diffuse interface
model if the dependence on $w$ is omitted.
The analogous observation for ``Model H'' has been already discussed in
\cite{GurtinTwoPhase}.
\item It is also possible to derive (\ref{sol1})-(\ref{sol6}) with the
approaches
discussed in Section \ref{eq:Derivation}.
\item Taking the definitions of $\mu_\varphi$ and $\bf K$ into account, we
observe that $\bf K$ in fact does not depend on $\bu$.
\end{itemize}

\section{Sharp interface asymptotics}\label{sec:SharpInterface}

In this section we identify the sharp interface limit of the diffuse
interface model introduced in the preceding sections. We will use the
method of formally matched asymptotic expansions where asymptotic
expansions in bulk regions have to match with expansions in
interfacial regions. There are four different asymptotic limits of
interest. Two use a constant mobility and will either lead to a model
where  diffusion takes place through the bulk or to a model without any
diffusion through the bulk, see also Abels and R\"oger \cite{AR1} for
the case when the densities in the two fluids are the same. Two 
cases are based on a mobility which is zero when the phase field takes
the values $\pm 1$. In this case diffusion through the bulk is of course not
observed in the sharp interface limit. But depending
on the scaling we will either see surface diffusion along the
interface or not. 

\subsection{The governing equations}

As usual for phase field models we introduce a scaling for $f$ with
respect to a small length scale parameter $\varepsilon$ as follows 
\begin{equation*}
f(\varphi, \nabla\varphi) =
\frac{\hat{\sigma}\varepsilon}{2}|\nabla\varphi|^2 + \frac{\hat{\sigma}}{\varepsilon}\psi(\varphi)\,
\end{equation*}
where $\hat{\sigma}$ is a constant related to the surface energy density.
As in Section 3, we choose the difference of volume fractions as order
parameter and we consider the following system
\begin{align}
\label{eq:mom} 
 \partial_t(\rho \ve) + \Div (\rho \ve\otimes \ve) &+ \Div (\ve\otimes
\tfrac{\tilde{\rho}_1-\tilde{\rho}_2}2 m(\varphi)\nabla\mu) -\Div
(2\eta(\varphi)D\ve)+ \nabla p\\
&= -\hat{\sigma}\eps\Div ( \nabla
\varphi \otimes \nabla \varphi),\nonumber\\
\label{eq:div} \Div \bu &= 0\,,\\
\label{eq:diff} \partial_t\varphi+(\bu\cdot\nabla)\varphi
&=\Div(m_\varepsilon(\varphi)\nabla\mu)\,,\\
\label{eq:chem} \mu& =
\frac{\hat{\sigma}}{\varepsilon}\psi'(\varphi)-\hat{\sigma}\varepsilon\Delta\varphi
+\beta'(\varphi)w\,,\\
\label{eq:weq} \partial_t w+(\bu \cdot\nabla) w &=
\Div(K(\varphi)w\nabla(g'(w)+\beta(\varphi))).
\end{align}
 To simplify the notation we drop the 
$\varphi$ as index in the chemical potential. We
assume that 
\begin{itemize}
\item[$\bullet$] $\rho(\varphi) = \frac{\tilde{\rho}_2-\tilde{\rho}_1}{2}
  \varphi + \frac{\tilde{\rho}_1+\tilde{\rho}_2}{2}$,
\item[$\bullet$] $g$ is convex, 
\item[$\bullet$] $\beta(\varphi)$ is smooth with $\beta(1)=\beta_2$,
  $\beta(-1) = \beta_1$,
\item[$\bullet$] $\eta(\varphi)$ is smooth and positive with $\eta(1)=\eta_2$,
  $\eta(-1)=\eta_1$,
\item[$\bullet$] $\psi(\varphi)$ is a double-well potential such that $\psi(1)
  = \psi(-1) =0$ and $\psi(z)>0$ if $z\not\in\{1,-1\}$,
\item[$\bullet$] $K(\varphi)$ is smooth, positive and  such that $K(1) = K_2$, $K(-1) = K_1$.
\end{itemize}

For the mobility $m_\varepsilon$ we distinguish four cases:
$$
m_\varepsilon(\varphi) = \begin{cases}
m_0 &\mbox{case I}\,,\\
\varepsilon m_0 &\mbox{case II}\,,\\
\frac{m_1}{\varepsilon}(1-\varphi^2)_+ &\mbox{case III}\,,\\
 m_1(1-\varphi^2)_+ &\mbox{case IV}\\
\end{cases}
$$
where $m_0,m_1 >0$ are constants and $(.)_+$ is the positive part of the quantity in the
brackets. The total relevant energy in this
scaling is
\begin{equation*}
F_\varepsilon (\varphi,\bu,w) =\int_\Omega \left\{\frac{\rho(\varphi)}{2}
|\bu|^2+ \frac{\varepsilon\hat{\sigma}}{2}
|\nabla\varphi|^2+ \frac{\hat{\sigma}}{\varepsilon} \psi(\varphi)
+ g(w)+\beta(\varphi)w\right\}\sd x\,.
\end{equation*}
For a solution
$(\bu^\varepsilon,p^\varepsilon,\varphi^\varepsilon,\mu^\varepsilon,w^\varepsilon)$
of the system (\ref{eq:mom})-(\ref{eq:weq}) we perform formally
matched asymptotic expansions. It will turn out that the phase field
$\varphi^\varepsilon$ will change its values rapidly on a length
scale proportional to $\varepsilon$. For additional information on
asymptotic expansions for phase field equations we refer to
\cite{FP,GNS}. 

\subsection{Outer expansions}

We first expand the solution in outer regions away from the
interface. We assume an expansion of the form $\bu^\varepsilon =
\sum^\infty_{k=0} \varepsilon^k\bu_k$, $\varphi^\varepsilon =
\sum^\infty_{k=0} \varepsilon^k\varphi_k,\dots$. An expansion of
(\ref{eq:chem}) in outer regions gives to leading order
$\psi'(\varphi_0)=0$ and we obtain the stable solutions $\pm 1$. We
will denote by $\Omega^\pm$ the regions where $\varphi_0  = \pm
1$. In the cases II-IV the leading order expansion of the other equations are
straightforward. We obtain:
\begin{align}
\label{eq:momsi} \rho_i(\partial_t\bu_0+\Div (\bu_0\otimes\bu_0))-2\eta_i\Div D\bu_0
+ \nabla p_0 &=0\,,\\
\label{eq:divsi} \Div\bu_0 &=0\,,\\
\label{eq:weqsi} \partial_t w_0+(\bu_0\cdot\nabla) w_0
&=K_i\nabla\cdot(w_0\nabla g'(w_0))\,,
\end{align}
where  $i=1,2$ for $x\in\Omega_-,\Omega_+$ and $\rho_1 =
\rho (-1)=\tilde \rho_1,
\rho_2 =
\rho (1)=\tilde \rho_2$. Due to the divergence free velocity we
obtain $2\Div D\bu_0=\Delta\bu_0$ and hence (\ref{eq:momsi})
simplifies to 
\begin{equation*}
\rho_i(\partial_t\bu_0+\Div(\bu_0\otimes\bu_0))-\eta_i\Delta\bu_0+\nabla
p_0=0\,.
\end{equation*} We remark that
(\ref{eq:weqsi}) leads to the convection diffusion equation 
\begin{equation*}
\partial_t w_0+(\bu_0\cdot\nabla) w_0=K_i\Delta w_0
\end{equation*}
in the case that $g(w)=w(\log w-1)$. In the cases II-IV we will not
need the chemical potential $\mu$ in the bulk. 

 In case I the flux
term $-m(\varphi)\nabla\mu$ will enter the momentum balance
(\ref{eq:mom}) to leading order in the bulk. 
In addition also the evolution equation 
(\ref{eq:chem}) for the phase field will contribute to leading order
to the sharp interface limit in the bulk. In fact in case I we
obtain  (\ref{eq:divsi}) and (\ref{eq:weqsi}) together with  
\begin{eqnarray*}
\tilde{\rho}_i(\partial_t\bu+\Div(\bu\otimes\bu))+\Div(\bu\otimes(\tfrac{\tilde{\rho}_1-\tilde{\rho}_2}{2})m_0\nabla\mu)-\eta_i\Delta
\bu +\nabla p&=&0\,,\\
\Delta\mu &=&0\,.
\end{eqnarray*}

\subsection{Inner expansions}\label{inner}

We now make an expansion in an interfacial region where a transition
between two phases takes place.

\subsubsection{New coordinates in the inner region}
We denote by $\Gamma =
(\Gamma(t))_{t\ge 0}$ the smoothly evolving interface which we expect to be 
the limit of the zero level sets of $\varphi$ when $\varepsilon$ tends to
zero and will now introduce new
coordinates in a neighborhood of $\Gamma$. Choosing a time interval
$I\subset\mathbb{R}$ and a spatial parameter domain
$U\subset\mathbb{R}^{d-1}$ we define a local parameterization 
\begin{equation*}
\bfgamma : I\times U\to \mathbb{R}^d
\end{equation*} 
of $\Gamma$. By $\bfnu$ we denote the unit normal to $\Gamma(t)$ pointing into
phase $2$ (which is the phase related to $\varphi=1$). 
Close to $\bfgamma(I\times U)$ we consider the signed distance function
$d(t,x)$ of a point $x$ to $\Gamma^0(t)$ with $d(t,x)>0$ if $x\in \Omega^+(t)$.
We now
introduce a local parameterization of $I\times\mathbb{R}^d$ close to
$\bfgamma(I\times U)$ using the rescaled distance $z=\frac d
\varepsilon$ as follows
\begin{equation*}
G^\varepsilon(t,\bs,z) := (t,\bfgamma(t,\bs)+\varepsilon
z\bfnu(t,\bs))\,
\end{equation*}
where $\bs\in U\subset \mathbb{R}^{d-1}$.
 We denote by
\begin{equation*}
\mathcal{V} = \partial_t\bfgamma\cdot\bfnu
\end{equation*}
the (scalar) normal velocity and 
observe that the inverse function
$(t,\bs,z) (t,x) := (G^\varepsilon)^{-1} (t,x)$ fulfills
\begin{equation*}
\partial_t z=\tfrac{1}{\varepsilon} \partial_t d =
-\tfrac{1}{\varepsilon} \mathcal{V}\,.
\end{equation*}
To derive the last identity we used (2.6) and (2.20) of
\cite{DDE}. For a scalar function $b(t,x)$ we obtain for $\hat{b}$ defined in the
new coordinates via $\hat{b}(t,\bs(t,x),z(t,x))=b(t,x)$ the identity
\begin{equation*}
\frac{d}{dt} b (t,x)=\partial_t z\partial_z\hat{b} +\partial_t
\bs\cdot\nabs\hat{b} +\partial_t\hat{b} =
-\tfrac{1}{\varepsilon}\mathcal{V} \partial_z\hat{b}+\,\,\mbox{h.o.t.}\,
\end{equation*}
where h.o.t. stands for higher order terms.
With respect to the spatial variables we obtain, see Appendix,
\begin{equation}\label{nabla}
\nabla_x b = \nabla_{\Gamma_{\varepsilon z}} \hat{b}
+\tfrac{1}{\varepsilon} \partial_z \hat{b} \,\bfnu
\end{equation} 
where $\nabla_{\Gamma_{\varepsilon z}}$ is the surface gradient on
\begin{equation*}
\Gamma_{\varepsilon z}:=\{\bfgamma(\bs)+\varepsilon z\bfnu(\bs)\mid \bs\in
U\}\,
\end{equation*}
where here and in what follows we often omit the $t$-dependence.
For a vector quantity $\mathbf{j}(t,x)$ written in the new coordinates
via $\hat{\mathbf{j}} (t,\bs(t,x),z(t,x)) = \mathbf{j}(t,x)$ we obtain
\begin{equation}\label{div}
\nabla_x \cdot\mathbf{j} =
\Div_{\Gamma_{\varepsilon z}}\hat{\mathbf{j}}
+\tfrac{1}{\varepsilon} \partial_z \hat{\mathbf{j}} \cdot\bfnu,
\end{equation}
where 
$\Div_{\Gamma_{\varepsilon z}}\hat{\mathbf{j}}$ is the
divergence of $\hat{\mathbf{j}}$ on $\Gamma_{\varepsilon z}$. 
In the Appendix we compute
\begin{equation}\label{delta}
\Delta_x b = \Delta_{\Gamma_{\varepsilon z}}\hat{b}-\tfrac{1}{\varepsilon}(\kappa+\varepsilon
z|\mathcal{S}|^2)\partial_z\hat{b}+\tfrac{1}{\varepsilon^2}\partial_{zz}\hat{b}+\,\,\mbox{h.o.t.}\,,
\end{equation}
where $\kappa$ is the mean curvature (the sum of the principal curvatures)
and $|\mathcal{S}|$ is the spectral norm of the Weingarten map $\mathcal{S}$. In addition
we note that (see Appendix) 
\begin{eqnarray*}
\nabla_{\Gamma_{\varepsilon z}}\hat{b} (\bs,z) &=& \nabla_\Gamma
\hat{b}(\bs,z)+\,\,\mbox{h.o.t.}\,,\\
\Div_{\Gamma_{\varepsilon z}}  
\hat{\bf{j}}(\bs,z)&=&
\Div_{\Gamma}
\hat{{\bf j}}(\bs,z)+\,\,\mbox{h.o.t.}\,,\\
\Delta_{\Gamma_{\varepsilon z}}
  \hat{b}(\bs,z)&=&\Delta_\Gamma\hat{b}(\bs,z)+\,\,\mbox{h.o.t.}\,
\end{eqnarray*}
where $\nabla_\Gamma,
\Div_{\Gamma},\Delta_\Gamma$ are the
surface gradient, the surface  divergence and the 
surface Laplacian on $\Gamma$.

\subsubsection{Matching conditions}
We now assume an $\varepsilon$-series approximation of the unknown
functions $\varphi,\mu , \bu,p,w,\dots$ which in the inner variables we will
denote by $\Phi,M, {\bf V}, P, W,\dots$. Denoting by $\Phi_0+\varepsilon\Phi_1+\dots$ the inner expansion and
by $\varphi_0+\varepsilon\varphi_1+\dots$ the outer expansion of the
phase field  we
obtain the following matching conditions at $x=\bfgamma(\bs)$:
\begin{eqnarray}\label{match1}
\underset{z\to\pm\infty}{\lim} \Phi_0(z,s)&=&\varphi_0(x\pm)\,,\\
\label{match2}
\underset{z\to\pm\infty}{\lim} \partial_z\Phi_1(z,s)
&=&\nabla\varphi_0(x\pm)\cdot\bfnu
\end{eqnarray}
 where $\varphi_0(x\pm),\dots$ denotes the limit
$\underset{\delta\searrow 0}{\lim}\,\varphi_0(x\pm\delta\bfnu)$. In addition we
obtain that if $\Phi_1(z,s)=A_\pm (s)+B_\pm(s)z+{o}(1)$ as
$z\to\pm\infty$ the identities
\begin{equation}\label{match3}
A_\pm(s)=\varphi_1(x\pm),\,\, B_\pm(s)=\nabla\varphi_0(x\pm)\cdot\bfnu
\end{equation}
have to hold (see \cite{F2}, \cite{Garst}). 
Of course similar relations hold for the other functions
like $\bu,\mu,\dots$. 

\subsubsection{The equations to leading order}
\label{leadorder}
 Plugging the asymptotic expansions into
(\ref{eq:mom})-(\ref{eq:weq}) we ask that each individual coefficient
of a power in $\varepsilon$ vanishes. The equation (\ref{eq:chem})
gives to leading order $\frac{1}{\varepsilon}$:
\begin{equation}\label{Leadphi}
0=\partial_{zz}\Phi_0-\psi'(\Phi_0)
\end{equation}
and from (\ref{match1}) we obtain
\begin{equation}\label{Leadphib}
\Phi_0(z)\to\pm 1\quad\mbox{for}\quad z\to\pm\infty\,.
\end{equation}
We now choose the unique solution of (\ref{Leadphi}),
(\ref{Leadphib}) which fulfills 
\begin{equation*}
\Phi_0(0)=0\,.
\end{equation*}
We in particular obtain that $\Phi_0$ does not depend on $t$ and
$s$. Equation (\ref{eq:div}) gives to leading order
\begin{equation}\label{VLO}
\partial_z{\bf V}_0\cdot\bfnu = \partial_z({\bf V}_0\cdot\bfnu)=0\,.
\end{equation}
The matching condition requires that $({\bf V}_0\cdot\bfnu)(z)$ is
bounded. Hence
\begin{eqnarray*}
(\bu_0\cdot\bfnu_0)(x+)&=&\underset{z\to\infty}{\lim}({\bf V}\cdot \bfnu_0)(z)
= \underset{z\to-\infty}{\lim}
({\bf V}\cdot\bfnu_0)(z)=(\bu\cdot\bfnu_0)(x-)\,.
\end{eqnarray*}
This implies
\begin{equation*}
[\bu_0\cdot\bfnu]^+_-=0\,,
\end{equation*}
where $[u]^+_-(x)=u(x+)-u(x-)$ denotes the jump of a quantity at the
interface. 

The analysis of (\ref{eq:diff}) now depends on the ansatz for the
mobility. We have to distinguish between four cases.\\[1ex]
\emph{Case I : $m_\varepsilon(\varphi)=m_0\,.$}\\
At order $\frac{1}{\varepsilon^2}$ we obtain from (\ref{eq:diff})
\begin{equation*}
0 = \partial_z(m_0\partial_zM_0\bfnu)\cdot\bfnu
=m_0 \partial_{zz}M_0\,.
\end{equation*}
Matching implies that $M_0$ is bounded and hence $M_0$ is constant. In
particular we derive
\begin{equation*}
[\mu_0]^+_-=0\,.
\end{equation*}
\emph{Case II : $m_\varepsilon(\varphi)=\varepsilon m_0\,.$}\\
At order $\frac{1}{\varepsilon}$ we conclude from (\ref{eq:diff})
\begin{equation}\label{aa33}
-\mathcal{V}
\partial_z\Phi_0+(\bu_0\cdot\bfnu)\partial_z\Phi_0=\partial_z(m_0\partial_zM_0\bfnu)\cdot\bfnu=m_0\partial_{zz}M_0\,.
\end{equation}
We obtain from matching
\begin{equation*}
\partial_zM_0\to 0\quad\mbox{for}\quad z\to\pm\infty\,.
\end{equation*}
Integrating (\ref{aa33}) with respect to $z$ gives
\begin{equation*}
\mathcal{V} = \bu_0\cdot\bfnu\,.
\end{equation*}
In addition we get $\partial_{zz}M_0=0$ and hence $M_0$ does not
depend on $z$. \\[1ex]
\emph{Case III : $m_\varepsilon(\varphi)=\tfrac{m_1}{\varepsilon}
(1-\varphi^2)_+\,.$}\\
At order $\frac{1}{\varepsilon^3}$ we get
\begin{equation*}
0=\partial_z(m_1(1-\Phi^2_0)\partial_zM_0\bfnu)\cdot\bfnu=\partial_z(m_1(1-\Phi^2_0)\partial_zM_0)\,.
\end{equation*}
Matching implies
\begin{equation*}
m_1(1-\Phi^2_0)\partial_zM_0\to 0\quad\mbox{for}\quad z\to\pm\infty\,.
\end{equation*}
We hence obtain
\begin{equation*}
m_1(1-\Phi^2_0)\partial_zM_0\equiv 0,
\end{equation*}
which implies
\begin{equation*}
M_0=M_0(s,t)\,.
\end{equation*}
\emph{Case IV : $m_\varepsilon(\varphi)= m_1(1-\varphi^2)_+\,.$}\\
At order $\frac{1}{\varepsilon}$ we get
\begin{equation*}
-\mathcal{V}\partial_z\Phi_0+({\bf
  V_0})\cdot\bfnu\partial_z\Phi_0=\partial_z(m_1(1-\Phi^2_0)\partial_zM_0)
\end{equation*}
and hence combining arguments from the Cases II and III above we
obtain
\begin{equation*}
\mathcal{V} = \bu_0\cdot\bfnu\quad\mbox{and}\quad M_0=M_0(s,t)\,.
\end{equation*}
We now analyze the diffusion equation for the soluble species.
Equation (\ref{eq:weq}) gives to leading order
$\frac{1}{\varepsilon^2}$
\begin{equation}\label{wlo}
\partial_z(K(\Phi_0)W_0\partial_z(g'(W_0+\beta(\Phi_0)))=0\,.
\end{equation}
Matching to the outer solution gives that $g'(W_0)+\beta(\Phi_0)$ is
bounded. Multiplying (\ref{wlo}) by $g'(W_0)+\beta(\Phi_0)$,
integration and integration by parts gives
\begin{equation*}
\int^\infty_{-\infty}
K(\Phi_0)W_0|\partial_z(g'(W_0)+\beta(\Phi_0))|^2dz=0\,.
\end{equation*}
Assuming $W_0>0$ we obtain that
\begin{equation*}
g'(W_0)+\beta(\Phi_0)
\end{equation*}
does not depend on $z$. Hence
\begin{equation*}
[g'(w_0)+\beta(\varphi_0)]^+_-=0\,.
\end{equation*}
For the choice $g(w)=w \log w -w $ we obtain 
\begin{equation*}
\log w_0(x+)-\log w_0(x-)+\beta_2-\beta_1=0
\end{equation*}
which yields the Henry jump condition
\begin{equation*}
\frac{w_0(x-)}{w_0(x+)} = \exp (\beta_2-\beta_1)\,.
\end{equation*}

Applying (\ref{nabla}) for each component we obtain 
\begin{eqnarray*}
\nabla_x\bu&=&\tfrac{1}{\varepsilon}\partial_z{\bf V}\otimes\bfnu+
\nabla_{\Gamma_{\varepsilon z}}{\bf V}\,,\\
D_x\bu &=& \tfrac{1}{2} \tfrac{1}{\varepsilon}
(\partial_z{\bf V}\otimes\bfnu+\bfnu\otimes\partial_z{\bf
  V})+\tfrac{1}{2}(\nabla_{\Gamma_{\varepsilon z}}{\bf
  V}+(\nabla_{\Gamma_{\varepsilon z}}{\bf V})^\top)\,.
\end{eqnarray*}
Defining $\mathcal{E}({\bf A})=\tfrac{1}{2}({\bf A}+{\bf A}^\top)$ for a
quadratic matrix $\bf A$
we compute
\begin{eqnarray*}
\nabla_x\cdot(\eta(\varphi)D_x \bu) &=&
\tfrac{1}{\varepsilon^2}\partial_z(\eta(\Phi)\mathcal{E}
(\partial_z{\bf V}\otimes\bfnu))\bfnu
+\tfrac{1}{\varepsilon}\partial_z(\eta(\Phi)\mathcal{E}(\nabla_{\Gamma_{\varepsilon
    z}}{\bf V}))\bfnu\\ &&
+\tfrac{1}{\varepsilon}\nabla_{\Gamma_{\varepsilon
    z}}\cdot(\eta(\Phi)\mathcal{E}(\partial_z{\bf V}\otimes\bfnu))
+\nabla_{\Gamma_{\varepsilon
    z}}\cdot(\eta(\Phi)\mathcal{E}(\nabla_{\Gamma_{\varepsilon z}}{\bf
  V}))\\
&=&\tfrac{1}{\varepsilon^2}\partial_z(\eta(\Phi)\mathcal{E}(\partial_z{\bf
  V}\otimes\bfnu)\bfnu)
+\tfrac{1}{\varepsilon}\partial_z(\eta(\Phi)\mathcal{E}(\nabla_{\Gamma_{\varepsilon
    z}}{\bf V})\bfnu)\\
&&+\tfrac{1}{\varepsilon}\nabla_{\Gamma_{\varepsilon
    z}}\cdot(\eta(\Phi)\mathcal{E}(\partial_z{\bf V}\otimes\bfnu))
+\nabla_{\Gamma_{\varepsilon
    z}}\cdot(\eta(\Phi)\mathcal{E}(\nabla_{\Gamma_{\varepsilon z}}{\bf
  V}))\,,
\end{eqnarray*}
where we used $\partial_z\bfnu=0$. We conclude from (\ref{VLO}) 
\begin{equation*}
(\bfnu\otimes\partial_z{\bf V}_0)\bfnu=(\partial_z{\bf V}_0\cdot\bfnu)\bfnu=0\,.
\end{equation*}
The fact that $\Phi_0$ does not depend on $t$ and $s$ and
(\ref{nabla}) imply
\begin{equation*}
\nabla\varphi\otimes\nabla\varphi=\tfrac{1}{\varepsilon^2}(\partial_z\Phi_0)^2(\bfnu\otimes\bfnu)
+\tfrac{1}{\varepsilon}\partial_z\Phi_1\partial_z\Phi_0(\bfnu\otimes\bfnu)+\,\,\mbox{h.o.t.}
\end{equation*}
and since $(\nabla_\Gamma \bfnu) \bfnu =0 $ we get
\begin{equation*}
\varepsilon\nabla\cdot(\nabla\varphi\otimes\nabla\varphi)=\tfrac{1}{\varepsilon^2}\partial_z(\partial_z\Phi_0)^2\bfnu+\tfrac{1}{\varepsilon}(\partial_z\Phi_0)^2(\nabla_\Gamma\cdot\bfnu)\bfnu
+\tfrac{1}{\varepsilon}\partial_z(\partial_z\Phi_1\partial_z\Phi_0)\bfnu+\,\,\mbox{h.o.t.}
.
\end{equation*}
Since the chemical potential to leading order does not depend on $z$
we observe that the term 
$ \Div (\ve\otimes
 m(\varphi)\nabla\mu)$ gives no contribution to the order $\frac
1{\varepsilon^2}$.
Hence we obtain at the order $\frac{1}{\varepsilon^2}$ from the
momentum equation 
\begin{equation}\label{MEQ-2}
\hat{\sigma}\partial_z(\partial_z\Phi_0)^2\bfnu
+\partial_z(\eta(\Phi_0)\partial_z{\bf V}_0)=0\,.
\end{equation}
Multiplying (\ref{MEQ-2}) with $\bfnu$, taking $\partial_z\bfnu=0$ and
$\partial_z{\bf V}_0\cdot\bfnu=0$ into account gives 
\begin{equation*}
\hat{\sigma}\partial_z((\partial_z\Phi_0)^2)=0\,.
\end{equation*}
Hence (\ref{MEQ-2}) implies
\begin{equation}\label{MEQ-2a}
\partial_z(\eta(\Phi_0)\partial_z{\bf V}_0)=0\,.
\end{equation}
Matching implies that ${\bf V}_0(z)$ is bounded. This implies that
(\ref{MEQ-2a}) interpreted as an ODE in $z$ has only solutions ${\bf V}_0$
which are constant in $z$. This implies after matching
\begin{equation}
[\bu_0]^+_-=0\,.
\label{vcont}
\end{equation}

\subsubsection{The generalized Gibbs--Thomson equation}
The equation for the chemical potential gives to the order $\eps^0$
\begin{equation}\label{phi1}
M_0=\hat{\sigma}\psi''(\Phi_0)\Phi_1-\hat{\sigma}\partial_{zz}\Phi_1+\hat{\sigma}\partial_z\Phi_0\kappa
  +\beta'(\Phi_0)W_0\,.
\end{equation}
In order to be able to obtain a solution $\Phi_1$ from (\ref{phi1}) a
solvability condition has to hold. This solvability condition will
yield the generalized Gibbs--Thomson equation. We  multiply 
(\ref{phi1}) with
$\partial_z\Phi_0$, integrate with respect to $z$ and obtain (using
the facts that $M_0$ and $V_0$ do not depend on $z$):
\begin{eqnarray*}
2M_0&=&\hat{\sigma}\int^\infty_{-\infty}(\psi''(\Phi_0)\partial_z\Phi_0\Phi_1-\partial_{zz}\Phi_1\partial_z\Phi_0)dz\\
&&+ \hat{\sigma} \kappa\int^\infty_{-\infty}(\partial_z\Phi_0)^2dz
+\int^\infty_{-\infty}\partial_z\beta(\Phi_0)W_0 dz\,.
\end{eqnarray*}
Defining $c_0:=\int^\infty_{-\infty}(\partial_z\Phi_0)^2\sd z$
we obtain after integration by parts, using the fact that 
$\partial_z \Phi_0 (z), \partial_{zz} \Phi_0 (z)$ decay exponentially
for $|z|\rightarrow \infty$,
\begin{eqnarray*}
2M_0&=&\hat{\sigma}\int^\infty_{-\infty}\partial_z(\psi'(\Phi_0)-\partial_{zz}\Phi_0)\Phi_1dz
+\hat{\sigma}c_0\kappa
\\&&+\int^\infty_{-\infty}\partial_z(\beta
(\Phi_0)W_0+g(W_0))dz
-\int^\infty_{-\infty}(\beta(\Phi_0)+g'(W_0))\partial_zW_0dz\,.
\end{eqnarray*}
Since $\beta(\Phi_0)+g'(W_0)$ does not depend on $z$ and since
$\psi'(\Phi_0)-\partial_{zz}\Phi_0=0$, see (\ref{Leadphi}), we obtain 
\begin{equation*}
2 \mu_0=\sigma\kappa+[g(w_0)-g'(w_0)w_0]^+_-\,
\end{equation*}
where $\sigma:= c_0 \hat{\sigma}$. 
This identity has been derived for a general function $g$. In the case
$g(w)=w\log w-w$ we obtain
\begin{equation*}
2\mu_0=\sigma\kappa-[w_0]^+_-\,.
\end{equation*}

\subsubsection{Interfacial flux balance in the sharp interface limit}
\label{diffequ}

We now expand the equations (\ref{eq:diff}) and (\ref{eq:weq}) further in order
to obtain contributions of the diffusive fluxes in the interface. The result
will depend on the choice of the mobility. In the cases II and IV the
interfacial diffusive fluxes for $\varphi$ are scaled such that they do not
contribute to a limiting sharp interface problem. We hence consider
only the cases I and III. \\[1ex]
\emph{Case I : $m_\varepsilon(\varphi) = m_0$.} \\
At order $\frac{1}{\varepsilon}$ we deduce from (\ref{eq:diff}) 
\begin{equation}\label{Idiff}
(-\mathcal{V}+{\bf
  V}_0\cdot\bfnu)\partial_z\Phi_0=\partial_z(m_0\partial_zM_1)\,.
\end{equation}
Matching gives $\partial_zM_1\to\nabla\mu_0\cdot\bfnu$ for
$z\to\pm\infty$. Integrating (\ref{Idiff}) gives
\begin{equation}
2(-\mathcal{V}+\bu_0\cdot\bfnu) = m_0[\nabla\mu_0\cdot\bfnu]^+_-\,.
\label{RHmass}
\end{equation}
\emph{Case III : $m_\varepsilon(\varphi) =
\frac{m_1}{\varepsilon}(1-\varphi^2)_+$.}\\
We have up to order $\eps^0$ in the interfacial region (setting
$m(\varphi)=m_1(1-\varphi^2)$)
\begin{eqnarray*}
&&\nabla\cdot(m(\varphi)\nabla\mu) =
\tfrac{1}{\varepsilon^2}\partial_z(m(\Phi_0)\partial_zM_0) +\\
&&\tfrac{1}{\varepsilon}\partial_z(m(\Phi_0)\partial_zM_1+m'(\Phi_0)\Phi_1\partial_zM_0)+\tfrac{1}{\varepsilon}\nabla_\Gamma\cdot(m(\Phi_0)\partial_zM_0\bfnu)\\
&&+\partial_z(m(\Phi_0)\partial_zM_2+m'(\Phi_0)\Phi_1\partial_zM_1)+\nabla_\Gamma\cdot(m(\Phi_0)\partial_zM_1\bfnu)+\nabla_\Gamma\cdot(m(\Phi_0)\nabla_\Gamma
M_0)\,.
\end{eqnarray*}
Using $\partial_zM_0=0$ we obtain from
(\ref{eq:diff}) at order $\frac{1}{\varepsilon^2}$ 
\begin{equation*}
0 = \partial_z(m(\Phi_0)\partial_zM_1)\,.
\end{equation*}
Matching gives
\begin{equation*}
m(\Phi_0)\partial_zM_1\to 0\quad\mbox{for}\quad z\to\pm\infty\,.
\end{equation*}
Hence $m(\Phi_0)\partial_zM_1=0$ and
\begin{equation*}
M_1=M_1(s,t)\,.
\end{equation*}
At order $\frac{1}{\eps}$ we obtain from
(\ref{eq:diff}), using $\partial_zM_1=0$ and $\partial_zM_0=0$, 
\begin{equation}\label{SD}
-\mathcal{V}\partial_z\Phi_0+({\bf
  V}_0\cdot\bfnu)\partial_z\Phi_0=\partial_z(m(\Phi_0)\partial_zM_2)+\nabla_\Gamma\cdot(m(\Phi_0)\nabla_\Gamma
M_0)\,.
\end{equation}
Matching gives
\begin{equation*}
m(\Phi_0)\partial_zM_2\to 0\quad\mbox{for}\quad z\to\pm\infty\,.
\end{equation*}
Integrating (\ref{SD}) gives
\begin{equation*}
2(-\mathcal{V} +\bu_0
\cdot\bfnu)=\int^\infty_{-\infty}\nabla_\Gamma\cdot(m(\Phi_0)\nabla_\Gamma
M_0)dz\,.
\end{equation*}
Since $m(\Phi_0)$ does not depend on $s$, we obtain
$\nabla_\Gamma\cdot(m(\Phi_0)\nabla_\Gamma M_0)=m(\Phi_0)\Delta_\Gamma
M_0$. Altogether we deduced 
\begin{equation}
2(-\mathcal{V}+\bu_0\cdot\bfnu) = \hat{m} \Delta_\Gamma\mu_0 \label{surfdif}
\end{equation}
where $\hat{m}=\int^\infty_{-\infty}m(\Phi_0)dz$. 

Finally we deduce the flux balance for $w$ at the interface in the
limit $\varepsilon\to 0$. 
Equation (\ref{eq:weq}) gives to order $\frac{1}{\varepsilon}$ 
\begin{equation}\label{Wflux}
-\mathcal{V} \partial_zW_0+({\bf
  V}_0\cdot\bfnu)\partial_zW_0=\partial_z(K(\Phi_0)W_0\partial_z(g''(W_0)W_1+\beta'(\Phi_0)\Phi_1))\,.
\end{equation}
Matching gives
\begin{equation*}
\partial_z(g''(W_0)W_1+\beta'(\Phi_0)\Phi_1)\to\nabla(g'(w_0)+\beta(\pm
1))\cdot\bfnu
\end{equation*}
for $z\to\pm\infty$. Integrating (\ref{Wflux}) gives
\begin{equation*}
(\mathcal{V}-\bu_0\cdot\bfnu)[w_0]^+_- = -[Kw_0 \nabla g^\prime (w_0 )]^+_-\cdot\bfnu.
\end{equation*}
In the case $g(w)=w \log
w-w$ this identity reduces to
\begin{equation*}
(\mathcal{V}-\bu_0\cdot\bfnu)[w_0]^+_- = -[K\nabla w_0]^+_-\cdot\bfnu.
\end{equation*}

\subsubsection{The momentum balance in the sharp interface limit }

We now want to analyze the momentum equation to the next order. The
term $\nabla\cdot(\eta(\varphi)D\bu)$ gives to the order
$\frac{1}{\varepsilon}$,
\begin{equation*}
\partial_z(\eta(\Phi_0)\mathcal{E}(\partial_z{\bf
  V}_1\otimes\bfnu)\bfnu)+\partial_z(\eta(\Phi_0)\mathcal{E}(\nabla_{\Gamma
    }{\bf V}_0)\bfnu)\,.
\end{equation*}
Matching requires
$\underset{z\to\pm\infty}{\lim}\partial_z{\bf V}_1(z)=\nabla\bu_0(x\pm)\bfnu$
and hence
\begin{equation}\label{matchv}
\partial_z{\bf V}_1\otimes\bfnu+\nabla_\Gamma{\bf
  V}_0\to\nabla_x\bu\quad\mbox{for}\quad z\to\pm\infty\,.
\end{equation}
In the cases II and IV the term $\Div 
 (\ve\otimes
 m_\varepsilon(\varphi)\nabla\mu)$ gives no contribution to order $\tfrac 1\varepsilon$
and hence
we obtain for the momentum equation at order
$\frac{1}{\varepsilon}$:
\begin{eqnarray*}
&-&\partial_z(\rho(\Phi_0){\bf
  V}_0)\mathcal{V}+\partial_z(\rho(\Phi_0)({\bf V}_0\otimes{\bf V}_0))\bfnu\\
&-&2\partial_z(\eta(\Phi_0)\mathcal{E}(\partial_z{\bf
  V}_1\otimes\bfnu)\bfnu)-2\partial_z(\eta(\Phi_0)\mathcal{E}(\nabla_{\Gamma
  }{\bf V}_0)\bfnu)
  \\
&-&\hat{\sigma}(\partial_z\Phi_0)^2\kappa\bfnu+\hat{\sigma}\partial_z(\partial_z\Phi_1\partial_z\Phi_0)\bfnu+\partial_zP_0\bfnu=0\,.
\end{eqnarray*}
Integrating with respect to $z$ gives after matching and using
(\ref{matchv})
\begin{equation*}
-[\rho_0\bu_0]^+_-\mathcal{V}
+[\rho_0\bu_0]^+_-\bu_0\cdot\bfnu-2[\eta\mathcal{E}(\nabla_x\bu_0)]^+_-\bfnu
-\hat{\sigma}\biggl(\int^\infty_{-\infty}(\partial_z\Phi_0)^2\sd
z\biggr)\kappa\bfnu+[p_0]^+_-\bfnu=0\,.
\end{equation*}
We obtain since $\bu_0\cdot\bfnu=\mathcal{V}$
\begin{equation}
-2[\eta
D\bu_0]^+_-\bfnu+[p_0]^+_-\bfnu= \sigma\kappa\bfnu\,.
\label{intstress1}
\end{equation}
\emph{Case I ($m_\varepsilon(\varphi)=m_0
$):} \\
We note that  $M_0$ does not depend on $z$ and that at $x=\bfgamma (s)$
$$\partial_z M_1\bfnu + \nabla_\Gamma M_0 \rightarrow 
\nabla \mu_0 (x\pm ) \qquad \mbox{for}\quad z\rightarrow \pm\infty.$$
For the momentum balance we hence get  to the order $\tfrac 1\varepsilon$
\begin{eqnarray*}
&-&\partial_z(\rho(\Phi_0){\bf
  V}_0)\mathcal{V}+\partial_z(\rho(\Phi_0)({\bf V}_0\otimes{\bf V}_0))\bfnu
-\partial_z ({\bf
  V_0} \otimes \tfrac 12 [\rho_0]_-^+ 
(\partial_z M_1\bfnu + \nabla_\Gamma M_0)) \bfnu  \\
&-&2\partial_z(\eta(\Phi_0)\mathcal{E}(\partial_z{\bf
  V}_1\otimes\bfnu)\bfnu)-2\partial_z(\eta(\Phi_0)\mathcal{E}(\nabla_{\Gamma
  }{\bf V}_0)\bfnu)
  \\
&-&\hat{\sigma}(\partial_z\Phi_0)^2\kappa\bfnu+\hat{\sigma}\partial_z(\partial_z\Phi_1\partial_z\Phi_0)\bfnu+\partial_zP_0\bfnu=0\,.
\end{eqnarray*}
Integrating this equation  with respect to $z$ we obtain since 
$$\int_{-\infty}^\infty \partial_z ({\bf
  V_0} \otimes \tfrac 12 [\rho_0]_-^+ 
(\partial_z M_1\bfnu + \nabla_\Gamma M_0)) dz = 
 \bu_0 \otimes  \tfrac 12 [\rho_0]_-^+ [\nabla \mu_0 ]_-^+ $$
the following interface condition
\begin{equation*}
[\rho_0]^+_-\bu_0(\bu_0\cdot\bfnu-\mathcal{V})-\tfrac 12 [\rho_0]^+_-
\bu_0 \, [\nabla \mu_0 ]^+_-\cdot\bfnu-[2\eta
D\bu_0]^+_-\bfnu+[p_0]^+_-\bfnu=\sigma\kappa\bu\,.
\end{equation*}
Using (\ref{RHmass}) we can simplify the above identity to
\begin{equation}\label{SII2}
-[2\eta D\bu_0]^+_-\bfnu+[p_0]^+_-\bfnu = \sigma\kappa\bfnu\,.
\end{equation}
We hence observe that in case I the sharp interface problem leads to the
classical conditions (\ref{vcont}), (\ref{SII2})
for $\bu$ and $p$ on the interface.

\emph{Case III ($m_\varepsilon(\varphi)=\tfrac{m_1}{\varepsilon}
(1-\varphi^2)_+$):} \\
Using $\partial_z M_0 =\partial_z M_1 =0$ we observe that 
$\Div
 (\ve\otimes
 m_\varepsilon  (\varphi)\nabla\mu)$
gives to the order $\tfrac 1\varepsilon$ the contributions
$$
m_1\partial_z((1-(\Phi_0)^2)_+\partial_zM_2({\bf V}_0 \otimes\bfnu))
\cdot \bfnu
+\Div_\Gamma(m_1(1-\Phi_0^2)_+({\bf V}_0 \otimes\nabla_\Gamma
M_0)).
$$
Taking the flux terms into account we obtain from the momentum balance 
at order $\tfrac{1}{\varepsilon}$:
\begin{eqnarray*}
&-&\partial_z(\rho(\Phi_0){\bf
  V}_0)\mathcal{V}+\partial_z(\rho(\Phi_0)({\bf V}_0\otimes{\bf
  V}_0))\bfnu\\
&+&
\tfrac12[\rho_0]_-^+m_1\partial_z((1-(\Phi_0)^2)_+\partial_zM_2({\bf V}_0\otimes\bfnu))
\cdot \bfnu\\
&+&\tfrac12[\rho_0]_-^+\Div_\Gamma(m_1(1-\Phi_0^2)_+({\bf V}_0\otimes\nabla_\Gamma
M_0)))\\
&-&2\partial_z(\eta(\Phi_0)\mathcal{E}(\partial_z{\bf
  V}_1\otimes\bfnu)\bfnu)-2\partial_z(\eta(\Phi_0)\mathcal{E}(\nabla_{\Gamma
  }{\bf V}_0)\bfnu)
  \\
&-&\hat{\sigma}(\partial_z\Phi_0)^2\kappa\bfnu+\hat{\sigma}\partial_z(\partial_z\Phi_1\partial_z\Phi_0)\bfnu+\partial_zP_0\bfnu=0\,.
\end{eqnarray*}
Integrating the above identity with respect to $z$ gives after matching 
\begin{equation*}
[\rho_0\bu_0]^+_-(\bu_0\cdot\bfnu-\mathcal{V})-
\tfrac 12[\rho_0]^+_- \hat m 
\Div_\Gamma(\bu_0\otimes \nabla_\Gamma \mu_0 )-2[\eta
D\bu_0]^+_-\bfnu+[p_0]^+_-\bfnu= \sigma\kappa\bfnu\,.
\end{equation*}
Using (\ref{surfdif}) we can simplify the above identity to
\begin{equation}
\label{intstress2}
-\tfrac 12[\rho_0]^+_- \hat m ((\nabla_\Gamma \mu_0 )\cdot\nabla_\Gamma)\bu_0-2[\eta
D\bu_0]^+_-\bfnu+[p_0]^+_-\bfnu=\sigma\kappa\bfnu\,.
\end{equation}

\begin{rem}
The interfacial stress balances (\ref{intstress1}) and
(\ref{intstress2}) have a normal and a tangential part.
For  (\ref{intstress1}) we obtain the classical conditions
$$  
-2\bfnu \cdot [\eta
D\bu_0]^+_-\bfnu+[p_0]^+_-= \sigma\kappa \quad \mbox{and} \quad
\bftau \cdot [\eta
D\bu_0]^+_-\bfnu =0\,$$
where the last identity holds for all $\bftau \in \R^d$ with
$ \bftau \cdot \bfnu =0$ and hence states the continuity of the tangential
interfacial stress.

The interfacial stress balance (\ref{intstress2})
has the normal part
$$
-\tfrac 12[\rho_0]^+_- \hat m \,\bfnu\cdot  ((\nabla_\Gamma \mu_0)
\cdot\nabla_\Gamma)\bu_0-2\bfnu\cdot [\eta
D\bu_0]^+_-\bfnu+[p_0]^+_-=\sigma\kappa
$$
and the tangential part
$$
-\tfrac 12[\rho_0]^+_- \hat m \bftau \cdot( (\nabla_\Gamma \mu_0)
\cdot\nabla_\Gamma)\bu_0-2\bftau \cdot[\eta
D\bu_0]^+_-\bfnu=0\quad\mbox{for all tangential} \quad \bftau,
$$
i.e., in this case the stress  $[\eta
D\bu_0]^+_-\bfnu$ can be discontinuous also in  tangential directions.

\end{rem}

\section{Free energy inequalities for the sharp interface limit}\label{sec:Energy}
The sharp interface limit is now given as follows. We search for a smoothly
evolving hypersurface $\Gamma$ which for all $t\ge 0$ separates
$\Omega$ into open sets $\Omega_-(t)$ and $\Omega_+(t)$ and we look for
functions $\bu ,p,w$ which are all defined for $x\in\Omega$ and $t\ge
0$. In the cases I and III we seek in addition  a chemical potential
$\mu$ which in case I is defined on $\Omega \times (0,\infty)$ and for
case III the potential $\mu$ is defined on the interface $\Gamma$
only.

Assuming for $g$ the form $g(w)=w(\log w -1)$ we need to solve 
in the cases II--IV  the system
\begin{eqnarray*}
\rho_i(\partial_t\bu+\Div(\bu\otimes\bu))-\eta_i\Delta\bu+\nabla p &=&
0,\\
\Div\bu&=&0,\\
\partial_t w+\bu\cdot\nabla w &=& K_i\Delta w
\end{eqnarray*}
in the bulk regions $\Omega_-$ and $\Omega_+$. 
In case I the momentum balance has to be replaced by
\begin{equation}\label{stressbalI}
  {\rho}_i\partial_t\ve +\Div \left(\ve\otimes\left({\rho}_i\ve +
\tfrac{{\rho}_1-{\rho}_2}2 m_0 \nabla
\mu\right)\right)-\eta_i\Delta \ve +\nabla p=0
\end{equation}
and 
 in
addition  we need to solve
\begin{equation*}
\Delta\mu =0
\end{equation*}
in the bulk. On the interface $\Gamma$ we require
\begin{eqnarray*}
[\bu]^+_- &=& 0\,,\\
\frac{w_2}{w_1} &=& \exp(\beta_1-\beta_2),\\
(\mathcal{V}-\bu\cdot\bfnu)[w]^+_- &=& -[K\nabla w]^+_-\cdot\bfnu\,
\end{eqnarray*}
where $w_1= w(x-)$ and $w_2= w(x+)$.
All other conditions depend on which of the cases I-IV we consider. \\[1ex]
\emph{Case I ($m_\varepsilon(\varphi)=m_0$):} In this case we require on the
interface 
\begin{eqnarray}\label{MB}
-[2\eta D\bu]^+_-\bfnu +[p]^+_-\bfnu &=& \sigma\kappa\bfnu,\label{stressbal}\\
\label{fluxba}
2(\bu\cdot\bfnu-\mathcal{V}) &=& m_0 [\nabla\mu]^+_-\cdot\bfnu, \\
\label{MVEQ}
2\mu &=&\sigma\kappa
-[w]^+_-\,.\label{genGT}
\end{eqnarray}
Here, it is remarkable that the
condition (\ref{MB}) is not affected by the diffusional flux and that the interface is not passively
transported by the interface, i.e.
$\bu\cdot\bfnu\neq \mathcal{V}$ is possible, compare
Abels and R\"oger \cite{AR1}.
\\[1ex]
\emph{Cases II and IV :} 
\begin{eqnarray*}
-[2\eta D\bu]^+_-\cdot\bfnu + [p]^+_-\bfnu &=&\sigma\kappa\bfnu \,,\\
\mathcal{V} &=& \bu\cdot\bfnu\,,
\end{eqnarray*}
i.e., in this case we recover a standard free boundary problem for the
Navier-Stokes system which in addition is coupled to the flow of a
soluble species. \\[1ex]
\emph{Case III ($m_\varepsilon(\varphi)=\tfrac{m_1}{\varepsilon}
(1-\varphi^2)_+$):} \\
We require that 
\begin{eqnarray}
\label{intstress2I}
-\tfrac 12[\rho]^+_- \hat m ((\nabla_\Gamma \mu)
\cdot\nabla_\Gamma)\bu-2[\eta
D\bu]^+_-\bfnu+[p]^+_-\bfnu&=&\sigma\kappa\bfnu
\,,\\
\label{surfdifI}
2(\bu\cdot\bfnu-\mathcal{V}) &=& \hat{m}\Delta_\Gamma\mu 
\,,\\
2\mu &=&\sigma\kappa
-[w]^+_- \,. \label{genGT3}
\end{eqnarray}
In this case the diffusion of the two components is limited to the
interfacial region. In fact the well-known surface diffusion flow
$\mathcal{V} =- \Delta_\Gamma\kappa$, see \cite{Mullins, CEN, EG1,
  EG2}, is in the new model coupled to fluid flow.
In this case we observe that the surface flux enters the interfacial 
stress balance.

In all cases we require $\bu=0$, $\nabla w \cdot \bfnu =0$ on $\partial \Omega$
and in case I we in addition require
$\nabla \mu \cdot \bfnu =0$ on $\partial \Omega$. In order to derive a free energy inequality for the sharp interface limit we need the following
transport identities, for a proof see \cite{DDE}.

\begin{lem} (Transport identities).
Let $\Gamma=(\Gamma(t))_{t\ge 0}$ with $\Gamma(t)\subset \Omega$, for all $t\geq 0$, be a smooth evolving hypersurface
and let $f$ be a quantity which is smooth for all $t\ge 0$ and
$x\in\Omega\setminus\Gamma(t)$ and such that $[f]_-^+$ exists at $\Gamma$. Then it holds 
\begin{equation*}
\frac{d}{dt}\int_{\Gamma(t)} 1\dsm =
-\int_{\Gamma(t)}\kappa \mathcal{V}\dsm =
-\int_{\Gamma(t)}
(\kappa\bfnu)\cdot(\mathcal{V}\bfnu)\dsm
\end{equation*}
and 
\begin{equation*}
\frac{d}{dt}\int_\Omega fdx=\int_\Omega\partial_tf dx
-\int_{\Gamma(t)}[f]^+_- \mathcal{V}\dsm\,.
\end{equation*}
\end{lem}

We are now in a position to compute the  dissipation rate and in conclusion  derive a free
energy inequality. 

\begin{theorem} (Free energy inequality).

A sufficiently smooth solution of the sharp interface problem  with $\Gamma(t)\subset \Omega$, for all $t\geq 0$, fulfills 
\begin{equation*}
\frac{d}{dt}\left[\int_\Omega \left(\frac{\rho}{2}|\bu|^2+(g(w)+\beta
    w)\right)\sd x +\int_{\Gamma(t)}\sigma\dsm\right]
=-\mathcal{D} \le 0\, ,
\end{equation*}
provided the integrals are finite.
Here $\rho=\rho_1,\rho_2$ and $\beta=\beta_1,\beta_2$ in the two
phases.
The quantity $\mathcal{D}$ is given as
\begin{align*}
&\mbox{\rm Case I} &:&\quad \mathcal{D} = \int_\Omega
2\eta|D\bu|^2\sd x +\int_{\Omega} K\tfrac{1}{w}|\nabla w|^2\sd x +\int_\Omega m_0|\nabla\mu|^2\sd x \,,\\
&\mbox{\rm Case II and IV} &:&\quad \mathcal{D} = \int_\Omega
2\eta|D\bu|^2\sd x +\int_{\Omega}K\tfrac{1}{w}|\nabla w|^2\sd x \,,
\\
&\mbox{\rm Case III} &:&\quad \mathcal{D} = \int_\Omega
2\eta|D\bu|^2\sd x +\int_{\Omega}K\tfrac{1}{w}|\nabla w|^2\sd x
+\int_{\Gamma(t)} \hat{m}|\nabla_{\Gamma}\mu|^2\dsm \,.
\end{align*}
\end{theorem}

\noindent
\begin{proof} Let us first discuss the terms which have to be treated
in the same way in all four cases.
The total free energy of the soluble species fulfills (using $g(w)=w\log
w-w$)
\begin{equation*}
\frac{d}{dt}\int_\Omega(g(w)+\beta w)dx = \int_\Omega\partial_tw(\log
w+\beta)\sd x-\int_{\Gamma(t)}[g(w)+\beta w]^+_-\mathcal{V}\dsm\,.
\end{equation*}
Since $\partial_t w+(\bu\cdot\nabla) w = K\nabla\cdot(w\nabla\log w)$, we
obtain
\begin{eqnarray*}
\int_\Omega\partial_t w(\log w+\beta)\sd x &=& \int_\Omega(\log w
+\beta)(-(\bu\cdot\nabla) w + K\nabla\cdot(w\nabla\log w))\sd x\\
&=& -\int_\Omega(\bu\cdot\nabla)(g(w)+\beta w)\sd x-\int_\Omega Kw|\nabla\log
w|^2\sd x\\
&&- \int_{\Gamma(t)}[(w\log w+\beta w)K\nabla\log w]^+_-\cdot\bfnu
\dsm\\
&=& -\int_\Omega Kw|\nabla\log w|^2\sd x+\int_{\Gamma(t)}[g(w)+\beta
w]^+_-\bu\cdot\bfnu\dsm\\
&&-\int_{\Gamma(t)}[(\log w+\beta)K\nabla w]^+_-\cdot\bfnu\dsm\\
&=& -\int_{\Omega } Kw|\nabla\log w|^2\sd x+\int_{\Gamma(t)}[g(w)+\beta
w]^+_-\bu\cdot\bfnu\dsm\\
&&+\int_{\Gamma(t)}(\log
w+\beta)(\mathcal{V}-\bu\cdot\bfnu)[w]^+_-\dsm
\end{eqnarray*}
where we used $\Div \bu=0$ and the fact that Henry's law implies the
continuity of $\log w+\beta$.
In addition we have
\begin{equation*}
\frac{d}{dt}\int_{\Gamma(t)}\sigma\dsm =
-\int_{\Gamma(t)}\sigma\kappa\mathcal{V}\dsm\,.
\end{equation*}

We now give a proof for the cases I and III in detail and
  discuss the cases II and IV afterwards.

{\it Case I:} We introduce the abbreviation 
$\widetilde{\vc{J}} = -\tfrac{\tilde{\rho}_2-\tilde{\rho}_1}{2}
m(\varphi)\nabla \mu
$.
 For
  the kinetic energy we compute, using a transport identity and (\ref{stressbalI})
\begin{eqnarray*}
\lefteqn{\frac{d}{dt}\int_\Omega\rho\frac{|\bu|^2}{2}\sd x  =
\int_\Omega\rho\bu\cdot\partial_t\bu\sd x -\int_{\Gamma(t)}[\rho]^+_-\frac{|\bu|^2}{2}\mathcal{V}\dsm}\\
&=&\int_\Omega\bu\cdot(-\Div (\bu\otimes(\rho\bu +\widetilde{\vc{J}})+2\eta\Div D\bu-\nabla
p)\sd x -\int_{\Gamma(t)}[\rho]^+_-\frac{|\bu|^2}{2}\mathcal{V}
\dsm\,.
\end{eqnarray*}
Using $\Div\bu=0$ 
and $\Div\widetilde{\vc{J}}=0$ in $\Omega_\pm$ we
obtain
\begin{equation*}
\bu\cdot\Div(\bu\otimes(\rho\bu+\widetilde{\vc{J}}))=\tfrac{1}{2}\nabla|\bu|^2\cdot(\rho\bu+\widetilde{\vc{J}})\,.
\end{equation*}
 Integration by
parts on $\Omega_+$ and $\Omega_-$ now gives
\begin{eqnarray*}
\frac{d}{dt}\int_\Omega\rho\frac{|\bu|^2}{2}\sd x &=& 
-\int_\Omega
2\eta|D\bu|^2\sd x+
 \int_{\Gamma(t)}
\tfrac{1}{2}|\bu|^2([\rho]^+_-(\bu\cdot\bfnu-\mathcal{V})+[\widetilde{\vc{J}}]^+_-\cdot\bfnu)ds_x\\
&&+\int_{\Gamma(t)}\bu\cdot([p]^+_-\bfnu -[2\eta D\bu]^+_-\bfnu)ds_x\\
&=& -\int_\Omega2\eta|D\bu|^2 dx + \int_{\Gamma(t)}
\sigma\kappa\bu\cdot\bfnu ds_x\,.
\end{eqnarray*}
where we used the stress balance (\ref{stressbalI}),
the interface conditions (\ref{MB}), (\ref{fluxba})  and the fact that $\bu$ is
continuous. 

We compute using  integration by parts, (\ref{fluxba})
and $\Delta\mu=0$ in the bulk
\begin{eqnarray}\label{FE2}
\int_{\Gamma(t)} 2(\bu\cdot\bfnu-\mathcal{V})\mu \dsm&=&
\int_{\Gamma(t)}m_0 \mu[\nabla\mu]^+_-\cdot \bfnu\dsm\\
&=& -\int_\Omega \nabla\cdot(\mu m_0 \nabla \mu)\sd x  =
-\int_\Omega m_0 |\nabla\mu|^2\nonumber\sd x  \,.
\end{eqnarray}

Altogether we obtain using the generalized Gibbs--Thomson law
(\ref{genGT}) and the fact that 
$g(w)+\beta w-(\log w+\beta)w = -w$
\begin{eqnarray}\label{FE1}
\lefteqn{\frac{d}{dt}\left(\int_\Omega\left(\rho\frac{|\bu|^2}{2}+g(w)+\beta
    w\right)dx +\int_{\Gamma(t)}\sigma\dsm\right)}\\
&=& -\int_\Omega 2\eta|D\bu|^2\sd x -\int_\Omega K\tfrac{1}{w}|\nabla w|^2\sd x \nonumber\\
&&+\int_{\Gamma(t)}(\bu\cdot\bfnu-\mathcal{V})(\sigma\kappa
-[w]^+_-)\dsm\nonumber\\
&=& -\int_\Omega 2\eta|D\bu|^2\sd x -\int_\Omega K\tfrac{1}{w}|\nabla
w|^2\sd x - \int_\Omega m_0 |\nabla\mu|^2\nonumber\sd x\,.\nonumber
\end{eqnarray}
{\it Case III:} Using (\ref{intstress2I}), (\ref{surfdifI}),
(\ref{genGT3}),
 setting  $\widetilde{\vc{J}}_\Gamma= \frac{\tilde \rho_1 -\tilde \rho_2}2
\hat m \nabla_\Gamma \mu $ and using the divergence theorem 
on manifolds we compute
\begin{eqnarray*}
\frac{d}{dt}\int_\Omega\rho\frac{|\bu|^2}{2}dx &=&
-\int_\Omega
2\eta|D\bu|^2dx+\int_{\Gamma(t)}\tfrac{1}{2}|\bu|^2[\rho]^+_-(\bu\cdot\bfnu-\mathcal{V})ds_x\\
&&+\int_{\Gamma(t)}\bu\cdot([p]^+_-\bfnu-[2\eta D\bu]^+_-\bfnu)ds_x\\
&=&-\int_\Omega 2\eta|D\bu|^2dx\\
&&+\int_{\Gamma(t)}\bu\cdot([p]^+_-\bfnu-[2\eta
D\bu]^+_-\bfnu-\tfrac{1}{2}\bu\Div_\Gamma \widetilde{\vc{J}}_\Gamma)ds_x\\
&=& -\int_\Omega 2\eta|D\bu|^2dx
+\int_{\Gamma(t)}\bu\cdot([p]^+_-\bfnu-[2\eta
D\bu]^+_-\bfnu+(\widetilde{\vc{J}}_\Gamma\cdot\nabla_\Gamma)\bu)ds_x\\
&=& -\int_\Omega 2\eta|D\bu|^2dx + \int_{\Gamma(t)} \sigma\kappa
\bu\cdot\bfnu ds_x\,.
\end{eqnarray*}
We now obtain
\begin{eqnarray*}
\frac{d}{dt}&&\hskip -9mm\left[ \int_\Omega \left(\frac{\rho}{2}|\bu|^2+g(w)+\beta
    w\right)\sd x +\int_{\Gamma(t)}\sigma\dsm\right]
\\&=& 
-\int_\Omega 2\eta|D\bu|^2\sd x -\int_\Omega K\tfrac{1}{w}|\nabla w|^2\sd x
\nonumber\\
&&+\int_{\Gamma(t)}(\bu\cdot\bfnu-\mathcal{V})(\sigma\kappa
-[w]^+_-)\dsm\nonumber\\
&=&
-\int_\Omega 2\eta|D\bu|^2\sd x -\int_\Omega K\tfrac{1}{w}|\nabla w|^2\sd x
+\int_{\Gamma(t)}\hat{m}(\Delta_\Gamma \mu)\mu\dsm\,.
\nonumber
\end{eqnarray*}
Applying the divergence theorem on manifolds for the last  term now gives the
result in case III.

 In the cases II and IV we have $\mathcal{V} = \bu\cdot\bfnu$ and
the calculations simplify. In particular we do not need to treat the
term $\int_{\Gamma(t)}(\bu\cdot\bfnu-\mathcal{V})2\mu\dsm$ and diffusion
plays no role for the fluid flow.
\end{proof}

\begin{rem}
Using formal asymptotic expansions one can also derive a
sharp interface limit for the non frame invariant model 
(\ref{non1})-(\ref{non4}). In this case the flux term
involving the gradient of the chemical potential
$\nabla\mu$  does not enter
the momentum balance law (\ref{stressbalI}).
The interface law (\ref{MB}) has to be replaced by
$$
[\rho]^+_-\bu(\bu\cdot\bfnu-\mathcal{V})-[2\eta D\bu]^+_-\bfnu +[p]^+_-\bfnu
= \sigma\kappa\bfnu
$$
and the summand involving $\nabla_\Gamma \mu$ in (\ref{intstress2I}) drops out.
In addition the generalized Gibbs-Thomson 
laws (\ref{genGT}) and (\ref{genGT3}) now contain an additional term involving
the velocity and  are given as follows
$$
2\mu =\sigma\kappa-\tfrac{1}{2}|\bu|^2[\rho]^+_--[w]^+_-.
$$
We also remark that the free energy inequality in Theorem 5.2
also holds in this case.
We refer to \cite{ModifiedModelH1}  for details.
\end{rem}
\section*{Appendix} 
We use the notation of Section \ref{inner} and prove
the identities (\ref{nabla}), \ref{div}) and (\ref{delta}). Let
$(s_1,\dots,s_{d-1})\in U$ and $G^\varepsilon_x
(s,z)=\bfgamma(s)+\varepsilon z\bfnu(s)$, where here and in what
follows we omit the $t$-dependence. Then
\begin{equation*}
\partial_{s_1}\bfgamma +\varepsilon
z\partial_{s_1}\bfnu,\dots,\partial_{s_{d-1}}\bfgamma +\varepsilon
z \partial_{s_{d-1}}\bfnu, \, \varepsilon\bfnu
\end{equation*}
is  a basis of $\mathbb{R}^d$ locally around $\Gamma$. We define the
metric tensor in the new coordinates as follows 
\begin{eqnarray*}
g_{ij} &=& (\partial_{s_i}\bfgamma +\varepsilon
z\partial_{s_i}\bfnu)\cdot(\partial_{s_j}\bfgamma+\varepsilon
z\partial_{s_j}\bfnu) \quad i,j=1,\dots,d-1\,,\\
g_{id} &=& g_{di}=(\partial_{s_i}\bfgamma+\varepsilon
z\partial_{s_i}\bfnu)\cdot\varepsilon\bfnu=0\quad i=1,\dots,d-1\,,\\
g_{dd} &=& \varepsilon\bfnu\cdot\varepsilon\bfnu=\varepsilon^2
\end{eqnarray*}
where we used that $\partial_{s_i}\bfnu\cdot\bfnu =
\frac{1}{2}\partial_{s_i}|\bfnu|^2=0$. We set
\begin{equation*}
\mathcal{G}_{\varepsilon z} =
(g_{ij})_{i,j=1,\dots,d},\,\,\hat{\mathcal{G}}_{\varepsilon z} =
(g_{ij})_{i,j=1,\dots,d-1},\, (\mathcal{G}_{\varepsilon
  z})^{-1}=(g^{ij})_{i,j=1,\dots,d},(\hat{\mathcal{G}}_{\varepsilon
  z})^{-1}=(g^{ij})_{i,j=1,\dots,d-1}\,
\end{equation*}
and hence
$$\mathcal{G}_{\varepsilon z} = \left(
\begin{array}{lc}
&0\\
\hat{\mathcal{G}}_{\varepsilon z}&\vdots\\
&0\\
0\cdots 0 &\varepsilon^2
\end{array}
\right)\,, \,\,
(\mathcal{G}_{\varepsilon z})^{-1} = \left(
\begin{array}{lc}
&0\\
\hat{(\mathcal{G}}_{\varepsilon z})^{-1}&\vdots\\
&0\\
0\cdots 0 &\varepsilon^{-2}
\end{array}
\right)\,.$$
Denoting by $s_d$ the $z$-variable we have for a scalar function
$b(t,x)=\hat{b}(t,s(t,x),z(t,x))$
\begin{eqnarray*}
\nabla_x b &=& \sum^d_{i,j=1} g^{ij}\partial_{s_i}\hat{b}\,\partial_{s_j}
G^\varepsilon_x
=\sum^{d-1}_{i,j=1} g^{ij}\partial_{s_i}\hat{b}\,\partial_{s_j}
G^\varepsilon_x +\tfrac{1}{\varepsilon^2}\partial_z\hat{b}\,\partial_z
G^\varepsilon_x\\
&=&\nabla_{\Gamma_{\varepsilon z}} \hat b
+\tfrac{1}{\varepsilon}\partial_z \hat b\,\bfnu\,
\end{eqnarray*}
where we used the fact that $g_{id}=0$ for $i=1,\dots,d-1$.
Here $\nabla_{\Gamma_{\varepsilon z}} \hat b$ is the surface gradient 
$\nabla_{\Gamma_{\varepsilon z}} b_{|\Gamma_{\varepsilon z}} $ on
$\Gamma_{\varepsilon z} := \{\bfgamma(s) +\varepsilon z\bfnu(s)\mid
s\in U\}$. In addition we compute for a vector quantity
${\bf j} (t,x)=\hat {\bf j} (t,s(t,x),z(t,x))$ 
\begin{eqnarray}
\label{divj}
\Div_x {\bf j} &=& \sum^d_{i,j=1} g^{ij}\partial_{s_i}\hat{{\bf
    j}}\cdot\partial_{s_j} G^\varepsilon_x
= \sum^{d-1}_{i,j=1} g^{ij}\partial_{s_i}\hat{{\bf
    j}}\cdot\partial_{s_j}G^\varepsilon_x
+\tfrac{1}{\varepsilon^2}\partial_z\hat{{\bf j}}
\cdot\varepsilon\bfnu\\
&=& \Div_{\Gamma_{\varepsilon z}}  \hat{{\bf j}}
+\tfrac{1}{\varepsilon}\partial_z\hat{{\bf j}} \cdot\bfnu\,,
\nonumber
\end{eqnarray}
where $\Div_{\Gamma_{\varepsilon z}}\hat{{\bf j}}$ is the
divergence on $\Gamma_{\varepsilon z}$. We remark that
$\nabla_{\Gamma_{\varepsilon z}}\hat{b}\cdot\bfnu =0\,,$
as $\bfnu$ is normal to $\Gamma_{\varepsilon z}$. We hence obtain
\begin{equation*}
\partial_z(\nabla_{\Gamma_{\varepsilon z}}\hat{b}\cdot\bfnu)=0
\end{equation*}
and
\begin{equation*}
\partial_z(\nabla_{\Gamma_{\varepsilon z}}\hat{b})\cdot\bfnu +
\nabla_{\Gamma_{\varepsilon z}} b\cdot\partial_z\bfnu=0\,.
\end{equation*}
Since $\partial_z\bfnu=0$, we get
\begin{equation*}
(\partial_z \nabla_{\Gamma_{\varepsilon z}}\hat{b})\cdot\bfnu=0\,.
\end{equation*}
We now compute
\begin{eqnarray*}
\Delta_x b &=&
\Div_x(\nabla_xb)=\Div_x(\nabla_{\Gamma_{\varepsilon
z}}\hat{b}
+\tfrac{1}{\varepsilon}\partial_z\hat{b}\,\bfnu)\\
&=&\Div_{\Gamma_{\varepsilon z}}(\nabla_{\Gamma_{\varepsilon z}}\hat{b})+\tfrac{1}{\varepsilon}(\nabla_{\Gamma_{\varepsilon z}}\partial_z\hat{b})\cdot\bfnu\\
&&+\tfrac{1}{\varepsilon}\partial_z\hat{b}\nabla_{\Gamma_{\varepsilon z}}\cdot\bfnu+\tfrac{1}{\varepsilon}(\partial_z\nabla_{\Gamma_{\varepsilon z}}\hat{b})\cdot\bfnu
+\tfrac{1}{\varepsilon^2}\partial_{zz}\hat{b}\cdot\bfnu+\tfrac{1}{\varepsilon^2}\partial_z\hat{b}\,\partial_z\bfnu\cdot\bfnu\,.
\end{eqnarray*}
Because of
$(\nabla_{\Gamma_{\varepsilon z}}\partial_z\hat{b})\cdot\bfnu=0$,
$(\partial_z\nabla_{\Gamma_{\varepsilon z}}\hat{b})\cdot\bfnu=0$,
$\bfnu=\nabla_x d$, (\ref{divj})  and
$\partial_z\bfnu=0$, we obtain
\begin{equation*}
\Delta_xb=\Delta_{\Gamma_{\varepsilon z}}
\hat b+\tfrac{1}{\varepsilon}(\Delta_xd)\partial_z\hat
b+\tfrac{1}{\varepsilon^2}\partial_{zz}\hat b\,.
\end{equation*}
Introducing $g_{\varepsilon z} =\det \mathcal{G}_{\varepsilon z}$ we get 
\begin{equation*}
\Delta_{\Gamma_{\varepsilon z}} \hat b = \frac{1}{\sqrt{g_{\varepsilon z}}}
\sum^{d-1}_{i,j=1} \partial_{s_i} (\sqrt{g_{\varepsilon z}}
g^{ij}\partial_{s_j}\hat{b})\,.
\end{equation*}
Since
\begin{equation*}
g_{ij}=\partial_{s_i}\bfgamma\cdot\partial_{s_j}\bfgamma
+\,\,\mbox{h.o.t.}\,,\,\, i,j=1,\dots,d-1,
\end{equation*}
we derive
\begin{eqnarray*}
\nabla_{\Gamma_{\varepsilon z}}\hat{b}(s,z)&=&\nabla_\Gamma
\hat{b}(s,z)+\,\,\mbox{h.o.t.}\,,\\
\Div_{\Gamma_{\varepsilon z}}\hat{{\bf
    j}}(s,z) &=& \Div_\Gamma\hat{{\bf j}}
(s,z)+\,\,\mbox{h.o.t.}\,,\\
\Delta_{\Gamma_{\varepsilon z}}\hat{b}(s,z)&=&\Delta_\Gamma
\hat{b}(s,z)+\,\,\mbox{h.o.t.}\,
\end{eqnarray*}
where $\nabla_\Gamma, \Div_\Gamma$ and $\Delta_\Gamma$ are computed 
on $\Gamma_{\varepsilon z}$ with the
metric tensor $\mathcal{G}_0$. From Gilbarg and Trudinger \cite{GT} (Lemma 14.17) we get (denoting by 
$\kappa_i$ the principal curvatures)
\begin{eqnarray*}
\Delta_xd &=& \sum^{d-1}_{i=1} \frac{-\kappa_i}{1-\kappa_id} =
\sum^{d-1}_{i=1} \frac{-\kappa_i}{1-\varepsilon\kappa_iz}
= -\sum^{d-1}_{i=1}\kappa_i-\sum^{d-1}_{i=1}
\varepsilon\kappa^2_iz+\,\,\mbox{h.o.t.}\\
&=& -\kappa-\varepsilon z|{\bf \mathcal{S}}|^2+\,\,\mbox{h.o.t.}\,,
\end{eqnarray*}
where $\kappa$ is the mean curvature and $|{\bf \mathcal{S}}|$ is the spectral norm of
the Weingarten map~${\bf \mathcal{S}}$.

Hence we obtain 
\begin{equation*}
\Delta_xb=\Delta_\Gamma\hat{b}-\tfrac{1}{\varepsilon}(\kappa+\varepsilon
z|{\bf \mathcal{S}}|^2)\partial_z\hat{b}+\tfrac{1}{\varepsilon^2}\partial_{zz}
\hat b+\,\,\mbox{h.o.t.}\,.
\end{equation*}


\def\cprime{$'$}

\end{document}